\documentclass[]{interact}

\usepackage{epstopdf}
\usepackage[caption=false]{subfig}

\usepackage{hyperref}
\usepackage{float}

\usepackage[british]{babel}
\usepackage[utf8]{inputenc}
\usepackage{epstopdf}
\usepackage{csquotes}
\usepackage[
    style=apa,
    backend=biber,
    sortcites=true,
    sorting=nyt,
    url=false,
    doi=false,
    hyperref=true,
    backref=false,
]{biblatex}

\DeclareLanguageMapping{british}{british-apa}

\theoremstyle{plain}

\theoremstyle{definition}

\theoremstyle{remark}

\begin{document}


\title{Integrated GIS- and network-based framework for assessing urban critical infrastructure accessibility and resilience: the case of Hurricane Michael}

\author{
\name{Pavel O. Kiparisov\textsuperscript{a}\textsuperscript{b}\thanks{Correspondence email: kiparisov\_pavel@phd.ceu.edu; https://orcid.org/0000-0003-1223-7964} and Viktor V. Lagutov\textsuperscript{a}\textsuperscript{c}}
\affil{\textsuperscript{a}Department of Environmental Sciences and Policy, Central European University (CEU), Vienna, Austria;
\textsuperscript{b}Biodiversity and Natural Resources (BNR) Program, International Institute for Applied Systems Analysis (IIASA), Laxenburg, Austria;
\textsuperscript{c}Food and Agriculture Organization of the United Nations (FAO), Budapest, Hungary.}
}

\maketitle

\begin{abstract}
This study presents a framework for assessing urban critical infrastructure resilience during extreme events, such as hurricanes. The approach combines GIS and network analysis with open remote sensing data of the aftermath, vector data on infrastructure, and socio-demographic attributes of populations in affected areas. Using Panama City as an example case study, this paper quantifies hurricane impacts on residents and identifies vulnerable locations for urban planners' attention. Simulations demonstrate how implementing measures at identified weak points can improve system resilience. Comparing pre-hurricane conditions with the aftermath and several years later allows observing network property changes and assessing overall resilience improvements. Findings indicate that individuals over 65 in the studied settlement are more susceptible to disasters, while males in this age category face higher risks.
\end{abstract}



\section{Introduction}
\label{intro}

Coastal urban areas bear the brunt of negative impacts from climate change, including a surge in hazards such as cyclones, floods, storms, erosion, sea-level rise, and heatwaves. These disasters pose a challenge to the resilience of coastal cities by testing their capacity to withstand shocks, adapt, recover, and ultimately improve. In the U.S., the combined effects of climate change and coastal development are projected to result in hurricane-related economic losses increasing at a faster rate than the overall growth of the economy and the number of individuals residing in areas with a high risk of experiencing substantial damage from hurricanes is expected to increase \autocite{dinan2017}. Despite these disincentives, individuals still continue to be attracted to coastal regions due to personal reasons and some remaining economic incentives. Those who already reside near the coast face significant challenges when attempting to relocate due to pre-existing investments, social ties, and limited availability of affordable housing in alternative locations \autocite{moser2017}.

Since cities inevitably stand in the way of natural disasters, access to critical infrastructure is essential for managing crises and protecting populations before, during, and after extreme events that disrupt access to services by destroying infrastructure and blocking roads with floodwaters, debris, and falling objects. The adverse effects of such disasters impede individuals' capacity to access services and hamper emergency responders from timely reaching those who require assistance. In light of the escalating magnitudes and fluctuations of cyclone seasons \autocite{roca2023}, it is imperative to gain a comprehensive understanding of how hurricanes impact people's accessibility to essential services. According to scientific literature, there remains a lack of research focused on quantifying the levels of vulnerability and adaptive capacity of infrastructure and at-risk people in coastal cities \autocite{wannewitz2024}. Consequently, one of Florida's coastal settlements affected by Hurricane Michael was chosen as the study case.

This paper investigates the underlying determinants of natural disasters by integrating findings from urban studies, network analysis, and disaster risk reduction research, while also employing methods derived from Geographic Information Systems (GIS) to better understand these complex phenomena. Satellite and aerial remote sensing technologies are capable of providing a wealth of data necessary for identifying affected regions, evaluating post-disaster damages, and supplying inputs to predictive models that can forecast the susceptibility of both inland and coastal areas to catastrophic events \autocite{klemas2015}. These data are made public by space and other government agencies, as well as private companies. The OpenStreetMap platform offers data of acceptable quality on the location of essential services and the layout and organization of transportation networks. Utilizing these and other publicly available sources, the study measures the extent of damages sustained by cities impacted by catastrophes. Additionally, it is evaluated if there have been any observable demonstrations of overall system robustness and resilience in its post-recovery state. Finally, simulations aimed at identifying more vulnerable parts of the city were conducted so that urban actors can pay closer attention to them. The \textit{primary objective} of this approach is to introduce a straightforward and adaptable framework that can assist urban planners and policymakers in identifying neighborhoods with their demographic characteristics which are more likely to lose access to essential services during a catastrophe. This approach will help them prepare for the next emergency, as well as plan for the placement of new services or relocation of existing ones. Generated resilience metrics can be used to track or adjust the progress of a recovery process.

This research focuses on the capacity for recovery and adaptation in urban environments while emphasizing factors that contribute towards vulnerability, robustness and resilience. Vulnerability is typically understood as the degree of susceptibility to negative effects of disasters \autocite{fuchs2018}. If a system is \textit{vulnerable}, it can be easily damaged and lose some of its functionality, so vulnerability is ``the potential to suffer loss or harm'' \autocite{burton2018}. The transition from a vulnerable state to a \textit{robust} one means that the system is no longer as easy to break or damage; instead, it becomes more resistant to stress. Accumulated damage from a hazard, however, can still cause too much stress and move the entire system into a broken state---what matters is how much stress is required, and how quickly this transition happens \autocite{carlson2002}. The next level of system state quality is \textit{resilient}. Resilience is a concept with numerous definitions that vary widely; it is classically understood as the ability of a system to recover quickly from damage \autocite{burton2018}. In addition to being vulnerable, robust, and resilient, a system can be \textit{antifragile}. Not only does it withstand impact and recover quickly, but it also becomes stronger as a result of the stress it is subjected to \autocite{taleb2013, johnson2013}. It is crucial that better systems are designed with due consideration for these capacities, ensuring their efficient functionality \autocite{goble2017}. 

Critical infrastructure studied in this paper encompasses \textit{healthcare facilities} like hospitals and clinics, \textit{emergency services} such as police stations and fire stations, pharmacy outlets, designated shelters, and \textit{transportation systems} comprising roadways and rail networks. Urban infrastructure is essential to the functioning of a city and necessary for an effective emergency response, and risk reduction strategies should prioritize safeguarding this vital infrastructure to ensure that facilities, communication systems, energy networks, and transportation systems can continue providing services during times of crisis \autocite{sendai2015}. It is acknowledged that mobility within cities both before, during, and following disasters predominantly occurs via roadways \autocite{dong2020}. As such, maintaining road connectivity and ensuring accessibility on these routes is of utmost importance for effective emergency responses.

The study conducted by \cite{urlainis2022} provides an extensive review of major extreme events that have occurred over the past three decades, highlighting a discrepancy between perceived levels of critical infrastructure preparedness and actual risk exposure. Meanwhile, the study by \cite{mejia2021} presents a numerical model designed to predict flooding resulting from extreme weather events such as Hurricane María, with the aim of assessing potential impacts on communities and vital infrastructure. Their research identifies vulnerable infrastructure in future flood scenarios. \cite{lovett2022} show that residential property damage caused by Hurricane Michael was extensive, with over two-thirds of affected residents reporting high levels of destruction. Additionally, communication issues and a lack of proactive law enforcement during recovery efforts were perceived as problematic by the public, while many respondents expressed fear regarding driving on roads with downed power lines and dealing with looters \autocite{lovett2022}.
 
In current research, the impact of disasters on population subgroups stratified by sex and age. In general, literature suggests that during hurricanes, the aged and those with medical fragility are particularly susceptible to the adverse effects of catastrophes \autocite{heagele2019}. In addition, it is known that pregnant women, families with children \autocite{rivera2021}, and low-income individuals \autocite{ribot2017, kiparisov2023}, suffer disproportionate harm during catastrophic events. These populations are less likely to engage in self-protective actions both prior to, during, and following disasters. In \cite{paul2018}, authors examine hydrometeorological fatalities in Texas from 1959 to 2016 and finds that while the total number of fatalities has increased, the per capita rate has significantly decreased. The highest number of fatalities occur on transportation routes due to flooding, followed by heat-related deaths in permanent residences. Demographic trends show that males are approximately twice as likely to die as females, but this disparity has decreased over time; adults are at the highest risk overall, while children are most at risk from flooding and the elderly from heat-related deaths \autocite{paul2018}.

Current study also enhances existing knowledge by improving upon previous methodologies through testing various network measures in real-world scenarios. Network science that emerged from the graph theory allows to represent urban topology in a simplified abstract manner. Fundamental elements of a network (or a graph) are \textit{vertices} (also known as nodes or points) and \textit{edges} (also known as links and arcs).  Research on hazard-prone interconnected infrastructure involving network science primarily focuses on identifying vulnerabilities and enhancing robustness and resilience \autocite{grubesic2013typological, pinnaka2015modelling, zhou2017robustness, kermanshah2017, akbarzadeh2019, xiang2021, kiparisov2021, kiparisov2023, balakrishnan2024}.  The role of urban form in disaster resilience was examined in \autocite{sharifi2021}. The authors in \cite{sun2020resilience} review resilience metrics and measurement for transportation infrastructure.

Finally, current research employs probabilistic models to simulate potential outcomes resulting from future hurricanes, with a specific focus on Hurricane Michael's devastation patterns and patterns exceeding its power, exploring contingency scenarios that may arise in subsequent events. Furthermore, one type of treatment scenario is proposed for further analysis. Computer simulations are widely used in literature to aid in enhancing both our cognitive understanding and ability to mentally simulate complex scenarios that might otherwise prove counterintuitive or unpredictable \autocite{sterman2002}. A number of papers is dedicated to hurricane simulations 
\autocite{pistrika2010, dinan2017, tomiczek2017}. The authors in \cite{bilskie2016} examine how sea level rise affects storm surge inundation caused by tropical cyclones using a high-resolution model that covers the northern Gulf of Mexico. The findings have implications for coastal restoration and long-term sustainability.

This Section briefly covered literature, and the rest of the paper is organized as follows: Section \ref{methods} explains the methodological framework; Section \ref{case} explores the case study and draws the way forward; Section \ref{conclusion} summarizes main findings.
\section{Methods}
\label{methods}
\subsection{The framework for ex-post and ex-ante resilience assessment}
Without focusing on both hazard \textit{and} vulnerability, without which disaster---that is, harm to human systems---would not even occur, it is not possible to comprehend where to allocate resources to mitigate the likelihood of catastrophe. In essence, a natural disaster such as a hurricane does not necessarily constitute a catastrophic event unless it encounters societal infrastructures with inherent weaknesses. A prevalent focus on adaptation in literature limits us in understanding the cause \autocite{ribot2017}, which prompts us to initially address the questions of \textit{why} a disaster occurred prior to addressing questions such as \textit{what to do}, emphasizing the importance of understanding the \textit{root cause}. Identifying the cause may give a better chance at providing working solutions for vulnerability reduction \autocite{ribot2017}. In lower-income countries, effective mitigation strategies can lead to a hundredfold reduction in human losses, as was seen in Bangladesh when a Cyclone Sidr hit the coastline in 2007, resulting in a significantly smaller number of fatalities compared to two similar cyclones in the past---Cyclone Gorky (1991) and Cyclone Bhola (1970)---all thanks to hazard identification programs, preparedness of communities, evacuation planning, and more effective response  \autocite{ribot2017}. In high-income countries, with many measures already in place, the reduction of losses may not be as high, but there is always room for further improvement.

The methodological framework used in this paper is the continuation of the evolving research line \autocite{kiparisov2021, kiparisov2023} and is grounded in Alexander's conceptualization of fit as the absence of misfit \autocite{alexander1964}. Due to the inherent unpredictability of exact timing and extent of future disaster events, it is more prudent to prioritize preparedness rather than relying on forecasts alone \autocite{taleb2013, blecic2019antifragile}. This research focuses on identifying what falters or what may falter regarding the people's access to critical infrastructure during a hurricane event, rather than solely emphasizing a limited set of prescribed practices. In summary, the paper assesses the susceptibility of individuals' access to essential services and resources in the aftermath of Hurricane Michael in 2018, identifying its far-reaching implications for the entire system. Simulations of hurricane-like events that replicate Hurricane Michael's damage pattern or exceed its threshold of destruction are employed, incorporating an adaptation scenario for most affected residential streets---all to identify vulnerable locations and potential mitigation strategies. This framework combines two major quantitative methods---geographic information systems (GIS) and network analysis---and relies on various openly accessible satellite, vector, and sociodemographic data. 

\subsection{Data sources and types}
The quality of the data directly impacts the accuracy of the resulting analysis. To effectively quantify the consequences of hurricane damage on the network, high-resolution raster or detailed vector data that provide information about the structure and connectivity of urban systems as well as the occurrence of road blockage due to flooding, debris, or fallen trees are required. This enables accurate assessments of the impact of hurricane damage on transportation networks in cities, allowing for better preparedness and response efforts during such events.

The selection of the components of the urban system is based on review of literature and availability of open data. The studied urban systems are:

    \begin{enumerate}
    	\item \textit{Transportation system}
    	\begin{enumerate}
    		\setlength\itemsep{0em}
    		\item Road infrastructure
            \item Railways
    	\end{enumerate}
    	\item \textit{Emergency system}
    	\begin{enumerate}
    		\setlength\itemsep{0em}
    		\item Police stations
    		\item Hospitals
            \item Clinics
    		\item Fire stations
    		\item Shelters
            \item Pharmacies
    	\end{enumerate}
    	\item \textit{Social system}
    	\begin{enumerate}
    		\setlength\itemsep{0em}
    		\item Approximate population count
    		\item Population density, disaggregated by age and sex
    	\end{enumerate}
    \end{enumerate}

\begin{figure}[h!]
    \centering
    \includegraphics[width=.7\textwidth]{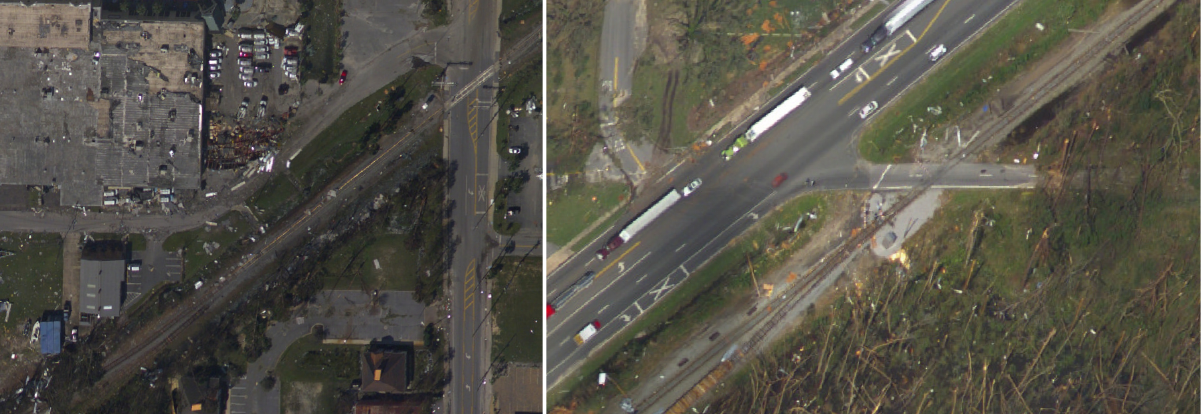}
    \includegraphics[width=.7\textwidth]{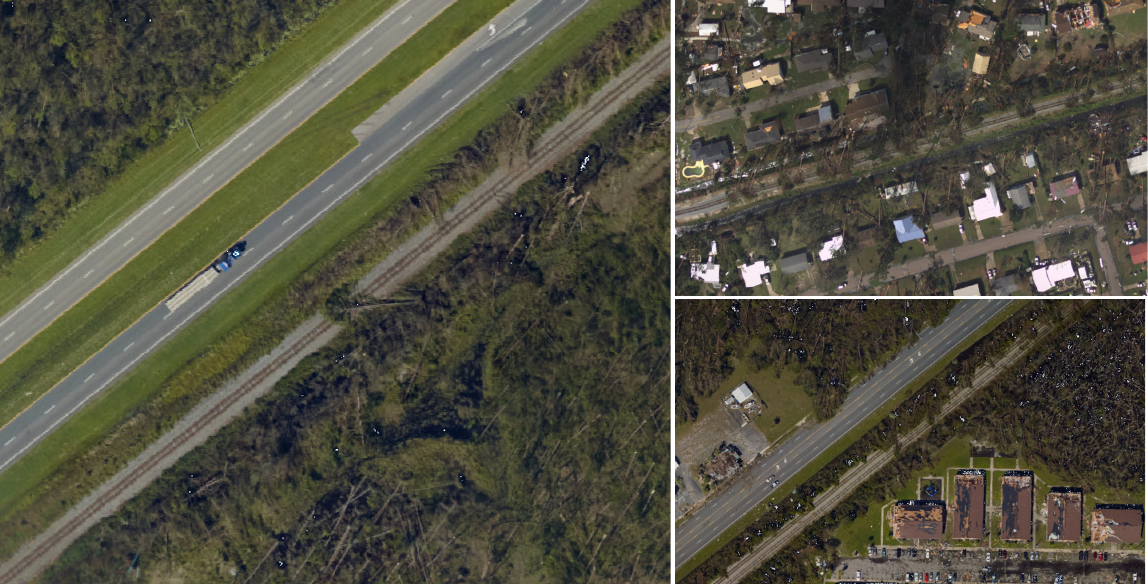}
    \caption{Fallen trees and debris blocking the railways. Source: \autocite{noaa2018}.} \label{fig:railwayblockage}
\end{figure}

For damage assessment, oblique and NADIR imagery that was captured following Hurricane Michael in October 2018 by the National Oceanic and Atmospheric Administration's (NOAA) Remote Sensing Division are used \autocite{noaa2018}. NOAA conducted aerial photography missions at altitudes ranging from 762 to 1524 meters, utilizing a Trimble Digital Sensor System (DSS) for data acquisition with the approximate ground sample distance (GSD) of each pixel being 25 cm. Samples of satellite imagery depicting road obstructions can be seen in \ref{fig:railwayblockage}. The fundamental structural information about the road network was obtained from a historical dataset of OpenStreetMap (OSM) \autocite{osm}. OSM is a project that unites volunteers who digitize physical objects on the surface. It should be noted, however, that both professional and amateur contributors provide input to the platforms, which raises legitimate questions regarding their reliability. Researchers and GIS professionals have evaluated the accuracy of OSM data, and their conclusion is that it is generally reliable for assessing urban resilience \autocite{herfort2015towards, sauter2019exploratory}.

\begin{table}[h]
    \footnotesize
	\centering
	\caption{Overview of data types and sources. The data related to transportation and facility location are obtained from \cite{osm}; road blockage data derived from the satellite imagery \autocite{noaa2018}; population density obtained from \cite{sedac2018a, sedac2018b}.}
	\begin{tabular}{l c r}
		\textbf{Data} &\textbf{Year} & \textbf{Source}\\
		\hline
		\textit{Physical} & & \\
		\hline
		Blockage & 2018 & NOAA \\
		Road & 2018, 2024 & OSM \\
        Railway & 2018, 2024 & OSM \\
		Fire station & 2018, 2024 & OSM \\
		Police station & 2018, 2024 & OSM \\
		Hospital  & 2018, 2024 & OSM \\
        Clinic  & 2018, 2024 & OSM \\
		Pharmacy  & 2018, 2024 & OSM \\
		Shelter  & 2018, 2024 & OSM \\
		\hline
		\textit{Social} & & \\
		\hline
		Total population count & 2015, 2020 & NASA \\
		Population density (age and sex) & 2010 & NASA\\
        \bottomrule
	\end{tabular}
	\label{tab:resources}
\end{table}

As the source of spatially embedded population data, a set of gridded datasets known as Gridded Population of the World, Version 4 (GPWv4) \autocite{doxsey2015} is utilized. The dataset titled "Population Count Adjusted to Match 2015 Revision of UN WPP Country Totals" \autocite{sedac2018a} for years 2015 (pre-disaster) and 2020 (post-disaster) was employed by us to estimate the number of persons per pixel in our study. The authors of this dataset ensured consistency with national censuses and population registers in terms of spatial distribution, and the figures were adjusted to align with the 2015 revision of the United Nation's World Population Prospects (UN WPP) country totals. To represent the demographic characteristics of human populations in the city, the paper employs the Basic Demographic Characteristics dataset \autocite{sedac2018b} depicting population density by sex and age. The estimates were derived by the authors of this dataset by calculating the proportions of males and females in each 5-year age group from ages 0-4 to ages 85+ based on national censuses and population registers. These proportions were then applied to the total population estimates for 2010 to obtain male and female populations by age. For current investigation, individuals are categorized into broader age groups based on sex. The first group comprises population density of males and females aged 0-14 years, the second group consists of males and females between 15 and 64 years old, and the third category includes individuals of males and females aged 65+. The data in both datasets were cropped from a global raster of a resolution of 30 arc-seconds (approximately 1 km at the equator).

\subsection{Data cleaning and preparation}

\begin{figure}[h]
    \centering
    \includegraphics[width=.37\textwidth]{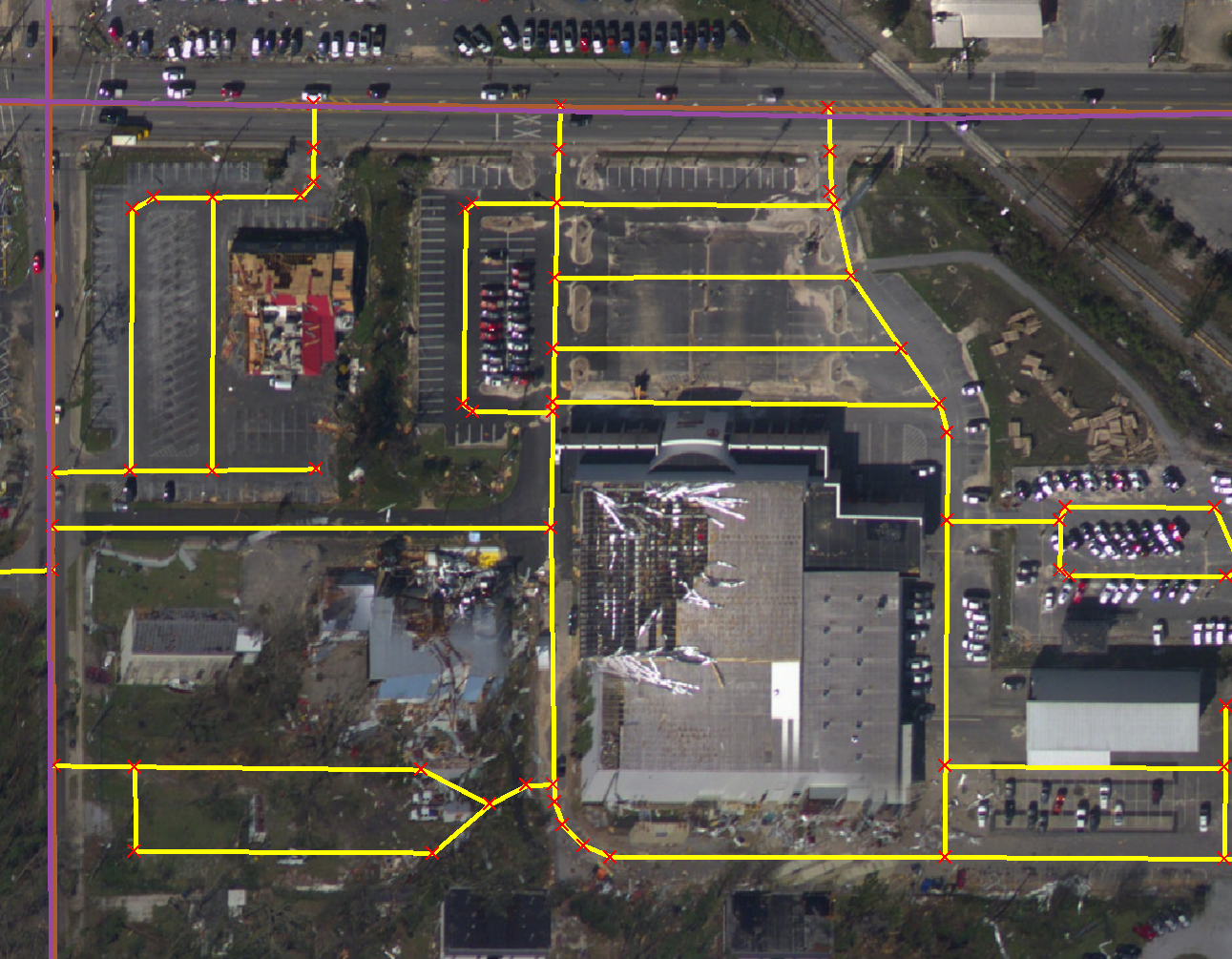}
    \caption{Missing digitization on the historical OSM data. In this example, yellow lines were missing in the 2018 historical vector data and were digitized manually to address the inaccuracies. Source:\cite{osm, noaa2018}.}\label{fig:missingdigitization}
\end{figure}

To effectively integrate the urban system's data structure into a GIS, it is necessary to perform a thorough preparatory stage of research. This involves cleaning and simplifying the street topography from OpenStreetMap, which often contains geometric errors. The street topology from the historical OSM-dump was compared to the de-facto satellite images and certain areas of the city should have been digitized manually to obtain accurate geospatial data for analysis (See Figure \ref{fig:missingdigitization}).

\begin{figure}[h!]
	\begin{center}
	\caption{The methodological workflow for the ex-post evaluation. Author's representation.}
	\includegraphics[width=14cm]{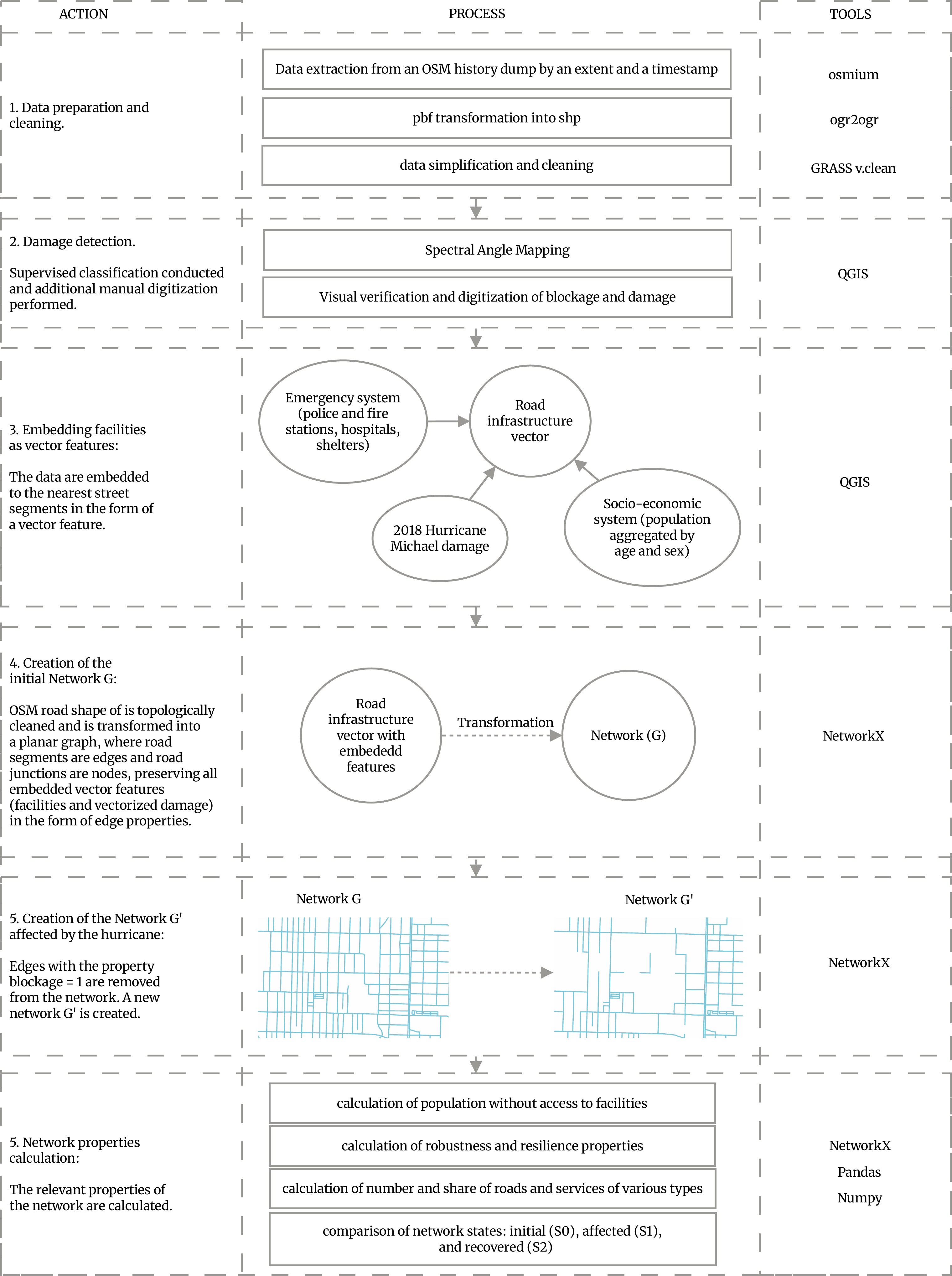}
	\label{tab:flowchart}
\end{center}
\end{figure}

Initially, each individual layer representing the transportation, emergency, and socio-economic aspects of the urban system must be created as separate data layers. Our task is to unify these individual layers into a single dataset that encompasses all components of the city's infrastructure. This consolidation is achieved by merging all elements on the road network as the common denominator. All elements are joined to the nearest road segment in the form of vector attributes, with each facility being connected to all the nearest edges rather than just one edge, in order to account for the possibility of accessing facilities from different sides of the building. For this reason, the framewor distinguishes between \textit{multiedge-facilities} and \textit{facility edges} \autocite{kiparisov2023}. The former is a single facility consisting of several edges that contain properties of a single facility; while the latter is a single edge that forms part of a multiedge-facility. In damage quantification, when a multiedge-facility loses all its edges, it is called \textit{loss of access}. If some but not all edges of the facility are blocked, his situation is defined as \textit{hindered access to services}. Therefore, road blockage becomes a boolean property of each road segment indicating $0$ when there is no obstruction and $1$ when the segment is obstructed by any type of barrier. Other GIS techniques such as fixing geometries, raster vectorization, and joining attributes by location are employed when necessary. The software used for this task is QGIS, a free and open-source desktop GIS application \autocite{qgis}. A flowchart outlining the complete workflow can be found in Figure \ref{tab:flowchart}.

\subsection{Geographic information systems (GIS)}

The initial classification was conducted using the Spectral Angle Mapper (SAM) algorithm. This algorithm is typically applied for satellite image classification using hyperspectral images, which provide detailed information about a surface's pixel spectrum and allow users to identify and distinguish between spectrally similar but unique materials \autocite{rashmi2014}. SAM is a supervised classification algorithm that identifies various classes in the image based on the calculation of the spectral angle. The spectral angle is calculated between a test vector constructed for each pixel and a reference vector built for each reference class selected by the user. Using SAM, thematic information is extracted---regions of interest (ROIs) such as clean roads, various debris, physical damage to roads, fallen tree blockage, and water occurrence on the roads (in most cases, the road blockages observed are primarily due to fallen trees).

\begin{figure}[h!]
    \centering
    \includegraphics[width=.355\textwidth]{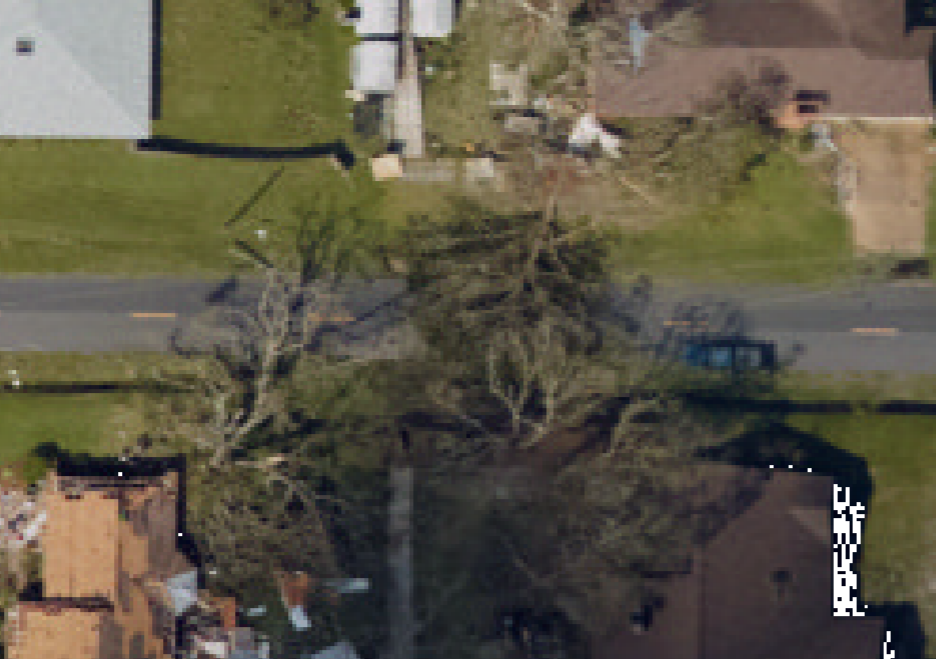}
    \includegraphics[width=.38\textwidth]{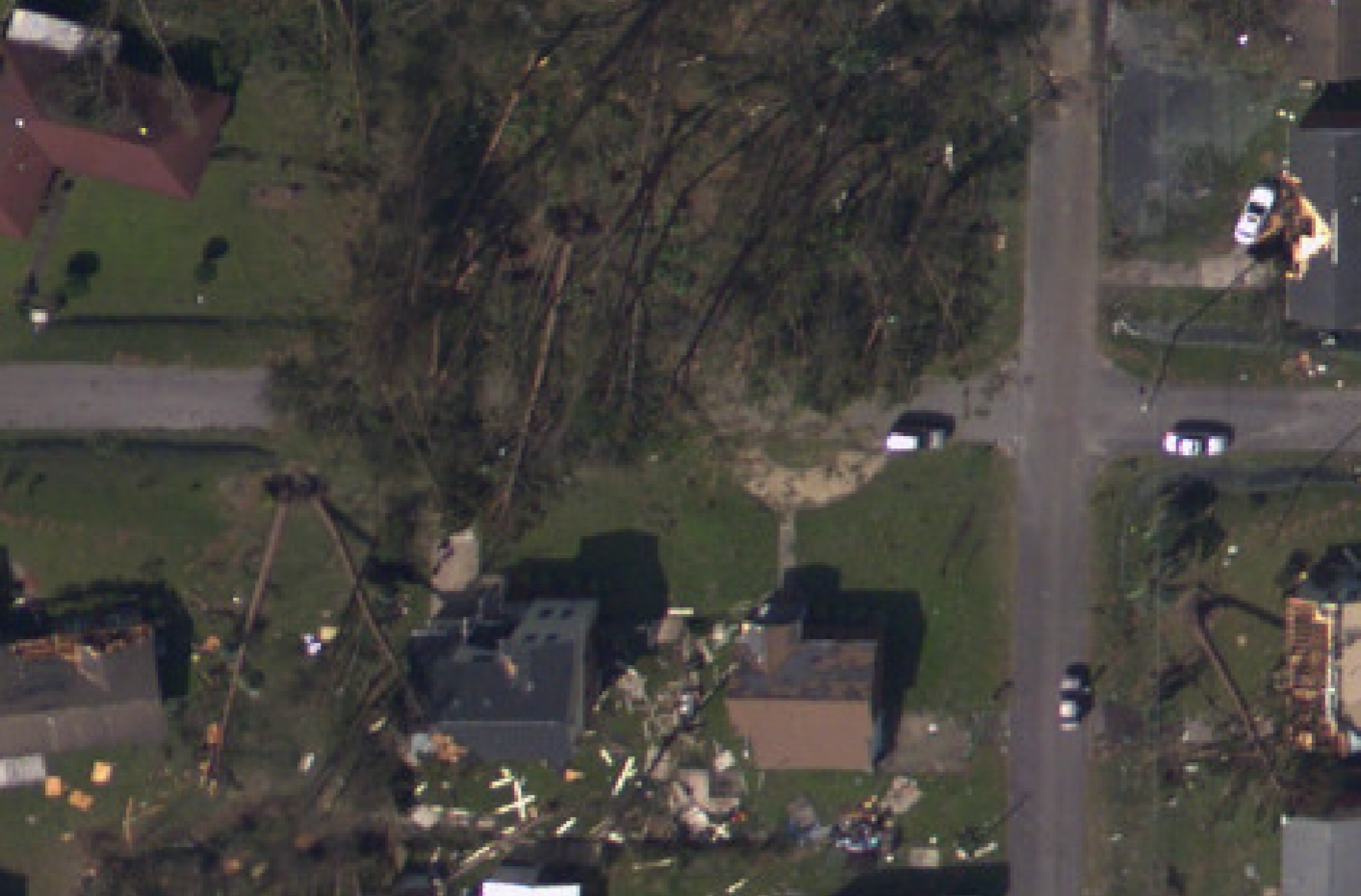}
    \caption{Fallen trees blocking the free passage of vehicles. Source: \cite{noaa2018}} \label{fig:treeblockage}
\end{figure}

\begin{figure}[h!]
    \centering
    \includegraphics[width=.95\textwidth]{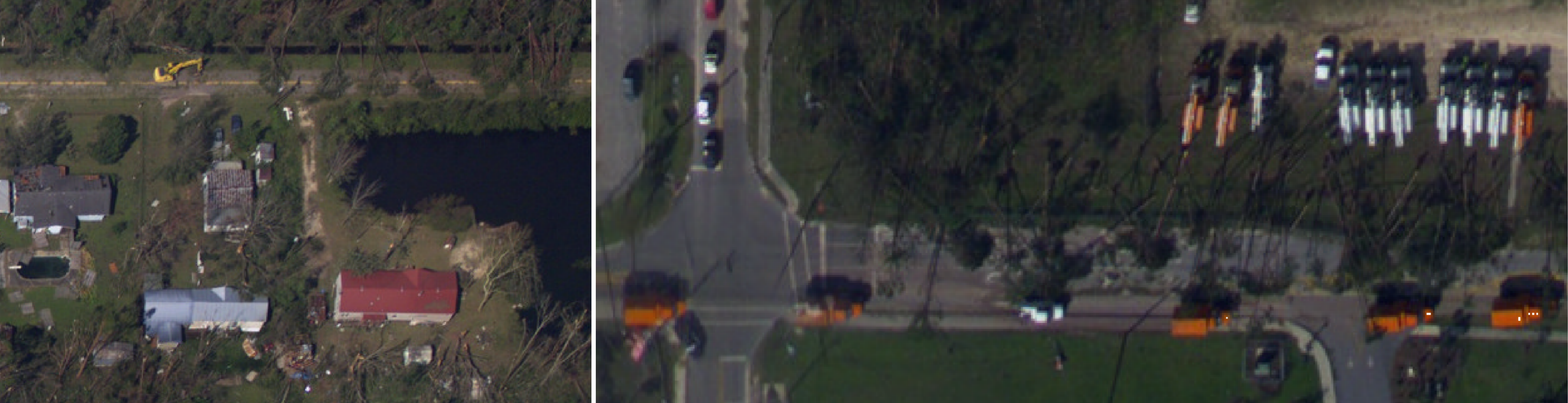}
    \centering
    \caption{On the left: An excavator cleans the road from fallen trees. On the right: Emergency vehicles are passing through the road blockage caused by the fallen trees. Source: \autocite{noaa2018}.}
    \label{fig:emergency}
\end{figure}

Visual verification is conducted on the entire study area to confirm that the registered instances of damage are accurate and reliable. While certain street segments are obviously blocked by fallen trees (See Figure \ref{fig:treeblockage}), in some cases it is possible to see that cars still drive on the side of the road (as depicted by emergency vehicles driving on the curb on the right-hand side of Figure \ref{fig:emergency}). These factors are considered when recording instances of damage. Obtaining access to ground-based survey data would enable more precise estimates of damages incurred, but the availability of data sources is limited, hence, accomplishing absolute precision in estimations remains difficult. A picture on the left-hand side of the Figure \ref{fig:emergency} shows a moment of cleaning up the road obstruction. Samples of railroad blockages are depicted in Figure \ref{fig:railwayblockage}.

\begin{figure}
    \centering
    \includegraphics[width=.37\textwidth]{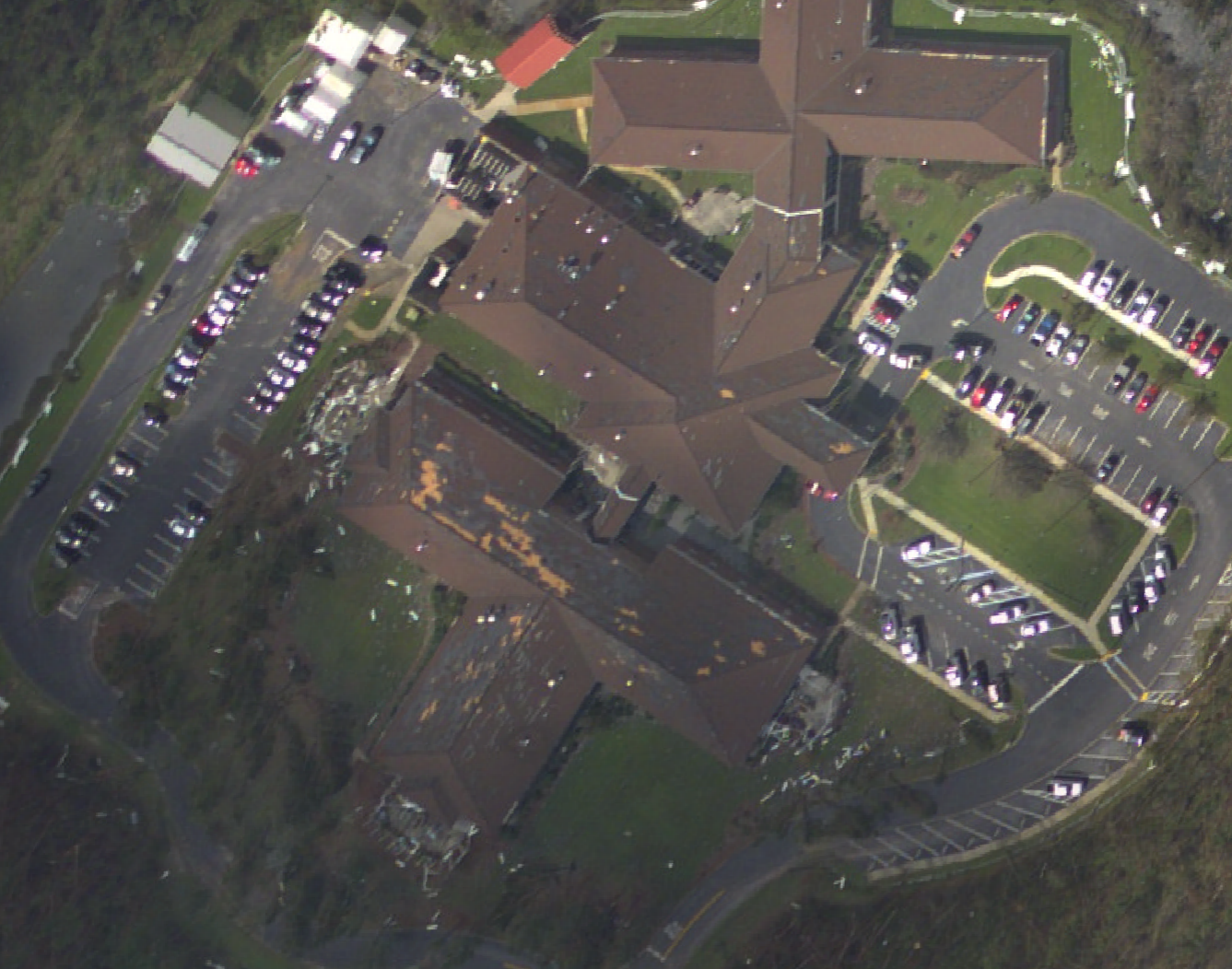}
    \centering
    \caption{An example of the facility damage to which was considered minor. Intact parking area and present accessibility indicate that this facility is fully operation despite visible roof damage. Source: \cite{noaa2018}.}
    \label{fig:minordamage}
\end{figure}

\begin{figure}[h!]
    \centering
    \includegraphics[width=.9\textwidth]{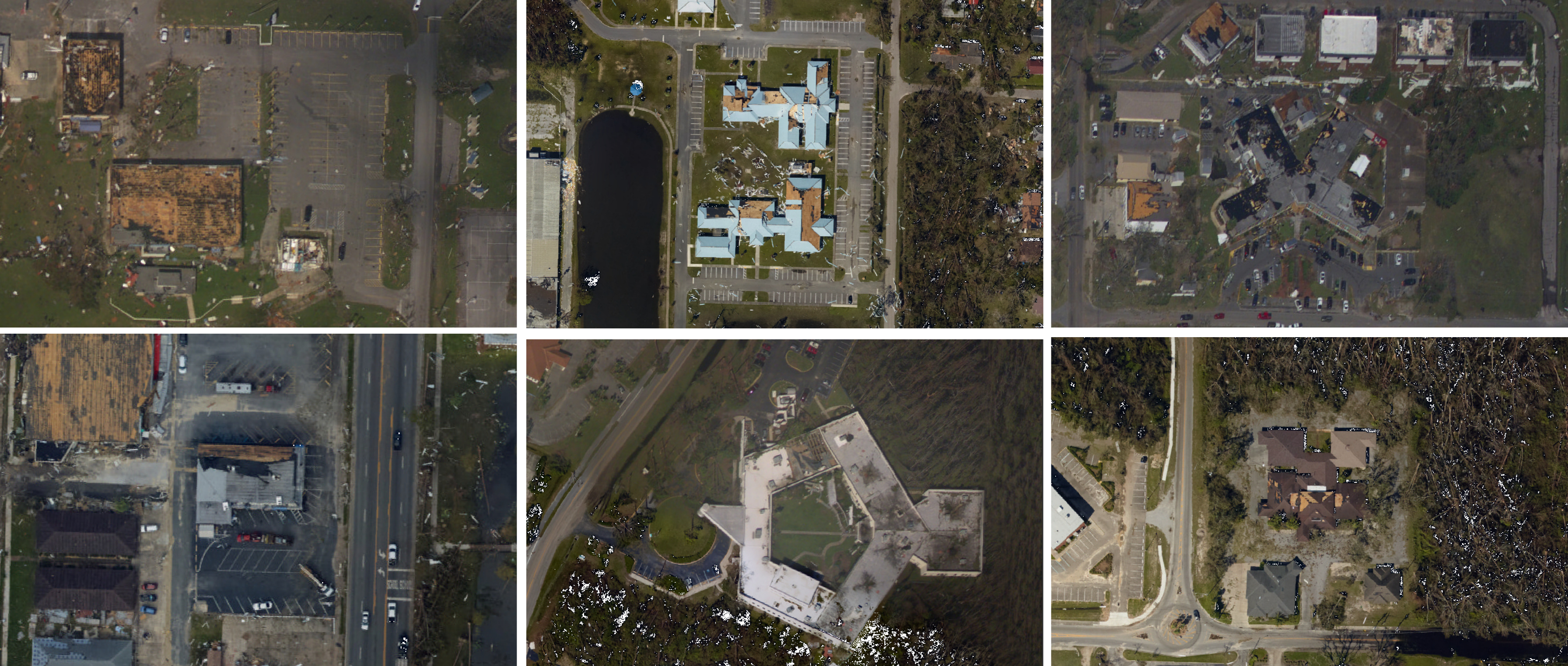}
    \caption{Examples of significantly damaged facilities (first column---fire station (top left) and pharmacy (bottom left), all the rest are hospitals and clinics. Source: \cite{noaa2018}.}
    \label{fig:damage}
\end{figure}

Another challenge was quantifying direct damages to critical facilities. Our visual assessment enabled us to identify which facilities appeared damaged but still operational (as seen in Figure \ref{fig:minordamage}), and which sustained too severe damage suggesting that they may not be able to provide services to the population (as depicted in Figure \ref{fig:damage}). Severely damaged facilities are considered as ``blocked'' due to low vehicle activity around them and significant visible damage. Furthermore, instances where emergency room facilities within damaged hospitals remained operational even though they had sustained damages are taken into accoount \autocite{abcnews2018}.

\subsection{Network description}
The cleaned geospatial data are then transformed into a two-dimensional graph (network) consisting of vertices and edges. This way, road infrastructure becomes a network of road segments (edges) that are connected by road junctions (vertices). Critical facilities, population density, and demographic characteristics are embedded in the network as edge properties.

The systems are defined as follows. System $S$, which represents an abstract representation of Panama City's streets, is composed of three forms - $S_0$, $S_1$, and $S_2$---that describe the system states before, during, and after the catastrophe respectively. Similarly, system $R$, which corresponds to an abstraction of the city's railways, also has three forms - $R_0$, $R_1$, and $R_2$---representing the pre-disaster state ($R_0$), the immediate post-catastrophe state ($R_1$), and several years into the recovery process ($R_2$). All subsystems are planar undirected graphs of the form $S_0 = (V_s, E_s)$ or $R_0 = (V_r, E_r)$, where $V_s$ is a set of vertices representing road (or railway in case of $V_r$) junctions (intersections) and $E_s$ is a set of edges representing parts of the road that link those junctions (railways lines linking junctions in case of $E_r$). Each edge $(u,v)$ in $E_s$ has a corresponding feature vector $\phi_{uv}$ indicating the elements of critical infrastructure which are adjacent to this edge, in particular, police station (equal to $0$ or $1$), hospital ($0$ or $1$), clinic, fire station ($0$ or $1$), shelter ($0$ or $1$), population (in count), and density of population of certain gender and age. $V_d$ and $E_d$ are the sets of vertices and edges affected by the damage from hurricane. Let $S_1= (V'_s, E'_s)$ and $R_1= (V'_r, E'_r)$ be copies of initial networks with all the edges from $E_d$ removed. Using Python \texttt{NetworkX} \autocite{hagberg2008}, basic network statistics and parameters are obtained and all the other calculations are done.

The calculations of \textit{number and share of roads and services of certain type} reveal what types of roads there are (e.g. trunk, primary, secondary, residential), what is the total share of affected edges of each type, how many and what critical services are available, what is the fracture of disabled services due to disasters. The computation of the \textit{percentage and average density of individuals categorized by sex and age who lost access to essential infrastructure} is presented as part of network description. This is calculated by looking at what is happening in \textit{subgraphs} of the network, which are the network's isolated lands. Essentially, it is checked if an isolated component still maintains access to at least one facility of each type. In literature, the component that maintains access to services is called \textit{robust component} \autocite{dong2020, kiparisov2023}. The population sitting in components that cannot reach a particular service is considered to be without access to the service. Evaluation of edge and node \textit{connectivity} was calculated along with network's \textit{distance} measurements, such as diameter and average shortest path length. Unlike \autocite{kiparisov2023}, this paper in addition to robustness discusses resilience because it now has a time perspective, therefore it looks not only at the immediate aftermath of the disaster but also observe how the system recovered over several years, hence a number of network measures were added to the analysis.

\begin{table}[h!]
    \caption{Summary of the key attributes related to robustness and resilience that were examined in the study.}
    \centering
    \footnotesize
    \begin{tabular}{p{2.5cm} p{3.5cm} p{6.6cm}}
    \toprule
       & Common formula & Description \\
    \midrule
    \textit{Average shortest path length} &  $L =\sum_{\substack{s,t \in V \\ s\neq t}} \frac{d(s, t)}{n(n-1)}$ &  The average shortest path length serves as an indicator of the connectivity and dispersion within a network or graph. It calculates the average distance that separates any two nodes in the network, considering all potential paths between them. \\
    \textit{Network efficiency} & $E = \frac{1}{N(N-1)} \sum_{i\neq j \in V} \frac{1}{d_{ij}}$ & Efficiency of a node pair in a graph refers to their multiplicative inverse, which represents the shortest path distance between them. In other words, it's the ratio of the number of efficient paths between two nodes and the total number of possible paths. The average global efficiency for all pairs of nodes in a graph is calculated by finding the weighted average of each node pair's efficiency across the entire network. \\
    \textit{Clustering} & $C = \frac{1}{n}\sum_{v \in G} c_v,$ & This measures how connected or clustered the entire network is by calculating the weighted average of each node's clustering coefficient across all nodes within the graph. \\
    \textit{Effective resistance} & $R = \sum_{1\leq i\leq j\leq N} E_{ij}$  & The effective graph resistance is calculated by finding the sum of the individual effective resistances for each pair of connected nodes across the entire graph. Overall, it measures how difficult it is to change the state of one node in a network while keeping all other nodes unchanged.\\
    \textit{Edge betweenness centrality} & ${b_e} =\sum_{s,t \in V} \frac{\sigma(s, t|e)}{\sigma(s, t)}$ & The betweenness centrality of an edge refers to the sum of the fraction of all possible shortest routes that traverse across it. \\
    \bottomrule
    \end{tabular}
    \label{tab:formulas}
\end{table}

\textit{Spectral measures}. Literature has proposed various spectral measures for describing the network robustness, including spectral radius \autocite{jamakovic2006, ellens2013}, vertex and edge connectivity \autocite{ellens2013, xu2023}, algebraic connectivity \autocite{jamakovic2008, ellens2013}, network efficiency \autocite{ellens2013, xiang2021, xu2023}, and effective resistance \autocite{ellens2013, yamashita2020}. The researchers have found that the following measures have a high predictability of the network robustness---effective resistance \autocite{yang2016, yamashita2020}, spectral radius \autocite{yamashita2020}, and natural connectivity \autocite{yamashita2020}. Algebraic connectivity can not be a good predictor as its measure is highly dependent on the topology of the graph \autocite{yamashita2020} and there is no statistical relationship with the network robustness \autocite{yang2016, yamashita2020}. Spectral radius, while being an essential attribute, is better suited for evaluating a system's robustness against the propagation of information or contagions \autocite{jamakovic2008, yamashita2020}, therefore this metric is not used in this research.

\textit{Clustering}. The clustering coefficient has been used to analyze network robustness \autocite{gao2019, zhu2022}, and it serves as an indicator of the local density and cohesion within a network, reflecting the extent to which connected nodes tend to cluster together. A higher clustering coefficient for a specific node suggests that its neighboring nodes are more likely to be interconnected, resulting in a more compact regional structure within the network.

\textit{Centrality}. The concept of edge betweenness centrality refers to the fracture of shortest paths that pass through an edge \autocite{eric2009}. Understanding the importance of edges can help us identify crucial roads that may be vulnerable during natural disasters. If a significant number of critical roads are at risk, it raises concerns about the overall system's resilience. On the other hand, if essential roads remain operational or alternative routes can efficiently take over their functions even in the event of major catastrophes, this may indicate that the network is robust and adaptable. Edge betweenness centrality is defined as in \cite{hagberg2008}.  The average and maximum of edge betweenness as a measure of network resilience is calculated. In essence, a reduced average and maximum value of centrality signifies greater resilience. In prior studies such as those by \cite{kermanshah2017} and \cite{akbarzadeh2019}, betweenness centrality has been utilized as a measure of network robustness, while \cite{xu2023} proposed a Gini index derived from node-level betweenness centrality to serve as an indicator of preparedness to a disaster.

\textit{System comparison}. After calculating all relevant statistics and collecting pertinent data, it is possible to compare three distinct states of our network: the initial configuration ($S_0 = (V_{S_0}, E_{S_0})$), the system impacted by the hurricane event ($S_1 = (V_{S_1}, E_{S_1})$), and its subsequent recovery state ($S_2 = (V_{S_2}, E_{S_2})$). The same applies to railway networks $R_0$, $R_1$, and $R_2$.

\subsection{Relationship identification} This involves performing simple regression analysis to identify relationships between the absence of access to critical services, an individual's age, and their sex and support or reject the hypothesis about certain population groups being more or less exposed to the disaster. 

\subsection{Simulation of the road damage}
\label{sec:simulation}
The results obtained from the probabilistic analysis allow us to create a spatial vulnerability map, which highlights areas that have experienced a loss of access to essential services due to road blockage. This is achieved by tracking all edges in the network subgraphs where at least one service has been disrupted after each simulation of the hurricane event. Although computer simulations provide merely a glimpse into possible outcomes in actual situations, they offer insights by illustrating probable consequences and guiding forthcoming strategies.

Two situations are examined, wherein a series of trials are conducted to test for mutually exclusive events ${A_1, A_2, A_3, ..., A_{300}}$ on the system $S_2$, forming an inherently stochastic process known as a \textit{Markov chain} \autocite{kolmogorov1963}:
\begin{enumerate}
    \item The initial scenario presents the implications brought about by Hurricane Michael with a certain likelihood of surpassing its destructive power. The blockage probability for each street type (trunk, primary, secondary, tertiary, residential, and other) is represented using a probability density function and by drawing a random number $\mathcal{E}$ for each street segment from the corresponding distribution. This task was done utilizing the \texttt{SciPy} Python library \autocite{virtanen2020}. The foundation parameters of the distributions are estimated based on Hurricane Michael historical data about road obstruction (as shown in Table \ref{tab:roads}); a likelihood of road blockage exceeding amounts caused by Hurricane Michael can be accounted for using a probability density function that follows the power law, which suggests that there is always a chance, albeit small, that an event will occur with, in our case, up to two times higher probability (Examples of distributions in Figure \ref{fig:pdfsample}). A probabilistic representation of the potential blockage for each individual street segment can be formulated as follows:
    \begin{equation}
        P(A_n^{trunk}) = P\{\mathcal{E} \in D\},
    \end{equation}
        where $D$ is a probability density function
    \begin{equation}
        \label{eq:prob}
        f(x,\alpha)=\alpha x^{\alpha-1}
    \end{equation}
     for $0.02\leq x \leq 0.04$, $\alpha=0.3$, and $\mathcal{E}$ is a random number drawn from the distribution generated by this function.
    \begin{equation}
        P(A_n^{primary}) = P\{\mathcal{E} \in D\},
    \end{equation}
        where $D$ represents the same function (\ref{eq:prob}) for $0.028\leq x \leq 0.056$, $\alpha=0.3$;
    \begin{equation}
        P(A_n^{secondary}) = P\{\mathcal{E} \in D\},
    \end{equation}
        where $D$ represents the function (\ref{eq:prob}) for $0.033\leq x \leq 0.066$, $\alpha=0.3$;
    \begin{equation}
    \label{eq:tertiary}
        P(A_n^{tertiary}) = P\{\mathcal{E} \in D\},
    \end{equation}
        where $D$ represents the function (\ref{eq:prob}) for $0.065\leq x \leq 0.13$, $\alpha=0.3$;
    \begin{equation}
        P(A_n^{residential}) = P\{\mathcal{E} \in D\},
    \end{equation}
        where $D$ is the function (\ref{eq:prob})  for $0.271\leq x \leq 0.542$, $\alpha=0.3$;
    \begin{equation}
        P(A_n^{other}) = P\{\mathcal{E} \in D\},
    \end{equation}
        where $D$ is the function (\ref{eq:prob}) for $0.06\leq x \leq 0.12$, $\alpha=0.3$.
    \item Given that a vast majority of street obstructions arise from fallen trees, it is reasonable to anticipate mitigating such occurrences at least as effectively as on tertiary streets, therefore our second scenario comprises an intervention plan aimed at enhancing resilience within residential areas specifically by amending the rules related to tree proximity along roadways and promoting only more hurricane-resistant species. Consequently, our previous calculations pertaining to the second scenario are revised using distribution function of a random variable for residential road segments comparable in robustness to those found on tertiary roads (parameters similar to equation \ref{eq:tertiary}):
    \begin{equation}
        P(A_n^{residential~treated}) = P\{\mathcal{E} \in D\},
    \end{equation}
        where $D$ represents the probability density function $f(x,\alpha)=\alpha x^{\alpha-1}$ \\ for $0.065\leq x \leq 0.13$, $\alpha=0.3$;
\end{enumerate}

Ultimately, the independent trial results are compiled into an aggregate score $v$, where $0\leq v \leq1$. This serves as a comprehensive measure of each location's susceptibility to damage. While this final outcome may not accurately reflect the precise probability of street destruction due to overestimation, it can be viewed as a ranking that highlights relative vulnerability and provides insight into which areas are more or less likely to experience failure. A global metric $V$ represents an average of the susceptibility levels to losing access to infrastructure ($v$) for every single road segment across the system.
\section{Case study}
\label{case}

\subsection{Background}
\label{background}
The US government once supported and subsidized shoreline development, which led to the establishment of numerous towns in hazard-prone areas. To address this issue, the Coastal Barrier Resources Act was passed in 1982, aiming to restrict federal expenditures and financial assistance that encouraged coastal development in high-risk areas \autocite{usgov1982}. Many communities, however, had already been well established by then. Currently, approximately 8.6 million individuals face an elevated risk of encountering flooding events with a probability of occurrence exceeding one percent \autocite{moser2017}. Furthermore, an additional 120 million citizens dwell in coastal regions vulnerable to storms similar to the one examined in this paper. A sobering estimate suggests that over two thousand US coastal towns and cities may be exposed to flooding and sea-level rise.  Significant portions of critical infrastructure remain uninsured, posing a substantial risk to both economic stability and public safety during extreme weather events such as hurricanes \autocite{moser2017}. Moreover, it has been reported that insurance companies have been withdrawing from high-risk areas prone to disasters \autocite{insurancebusinessmag2024}.

To provide context, the Florida's climate is humid, and its surface is predominantly low-lying and flat, with its highest elevation reaching just 114 meters above sea level  \autocite{hine1988}. Should sea levels rise only moderately, much of Florida's coastal Everglades would be flooded, potentially forming an open sea \autocite{hine2003}. Florida is also one of the most impacted states in the U.S. to hurricanes, with three out of five of the most devastating hurricanes in American history occurring within its borders \autocite{mcentire2011designing}. One such disaster was Hurricane Andrew, a category five storm that struck in 1992 and revealed significant shortcomings in Florida's approach to disaster management. At the time, the state's focus was primarily on preparedness, response, and recovery rather than mitigation \autocite{mcentire2011designing}. Addressing these shortcomings, the governor of Florida established the Disaster Planning and Response Review Committee, which generated ninety-four recommendations aimed at improving communication and coordination across all levels of government, bolstering evacuation plans, enhancing shelter capabilities, and providing more robust training \autocite{mcentire2011designing}. Despite these measures, however, the state's approach remained predominantly focused on post-disaster management and relief rather than promoting community resilience and fostering capacity for future disasters. The recommendations were implemented under the auspices of the Florida Department of Community Affairs through the Emergency Management, Preparedness, and Assistance Trust Fund \autocite{mcentire2011designing}.

In 1995, Florida began implementing mitigation programs, including a comprehensive planning process and the adoption of the Florida Building Code. As one of seven states required to address disaster issues in their comprehensive plans, Florida took significant steps towards promoting community resilience and reducing vulnerability to future disasters. Furthermore, changes made to the building code contributed significantly to the development of mitigation practices within the state \autocite{mcentire2011designing}. As a result, Hurricane Ivan, which struck Florida in 2004, revealed significant improvements in the state's disaster management system compared to the response to previous hurricanes. This was due in part to instrumental policy learning as well as social policy learning, as residents began to view natural disasters not as random events but rather as predictable and manageable threats that required proactive planning and preparation. Since Hurricane Andrew, communities have become increasingly aware of the importance of institutions and collective action in mitigating the impacts of disasters \autocite{mcentire2011designing}.

Hurricane Michael was a major test of Florida's resilience. In October 2018, Category Five Hurricane Michael caused significant damage across Central America and parts of the United States. The storm reached its peak intensity on October 10th with winds up to 260 km/h before making landfall near Mexico Beach, Florida---an area that serves as the primary focus for this study. The morning before landfall, sensors in Mexico Beach forecasted inundation equal to about 4.5 meters high, and the damage from Michael in the United States was estimated at about USD 25 billion \autocite{beven2019national}. Of this, approximately USD 18.4 billion occurred in Florida. The hazard was directly responsible for sixteen human lives. Seven deaths occurred in Florida, and three of them in or near Mexico Beach. The hurricane is also responsible for forty three indirect deaths (caused due to clean-up, traffic accidents, and medical issues) \autocite{beven2019national}.

\begin{figure}[h!]
    \centering
    \includegraphics[width=1\textwidth]{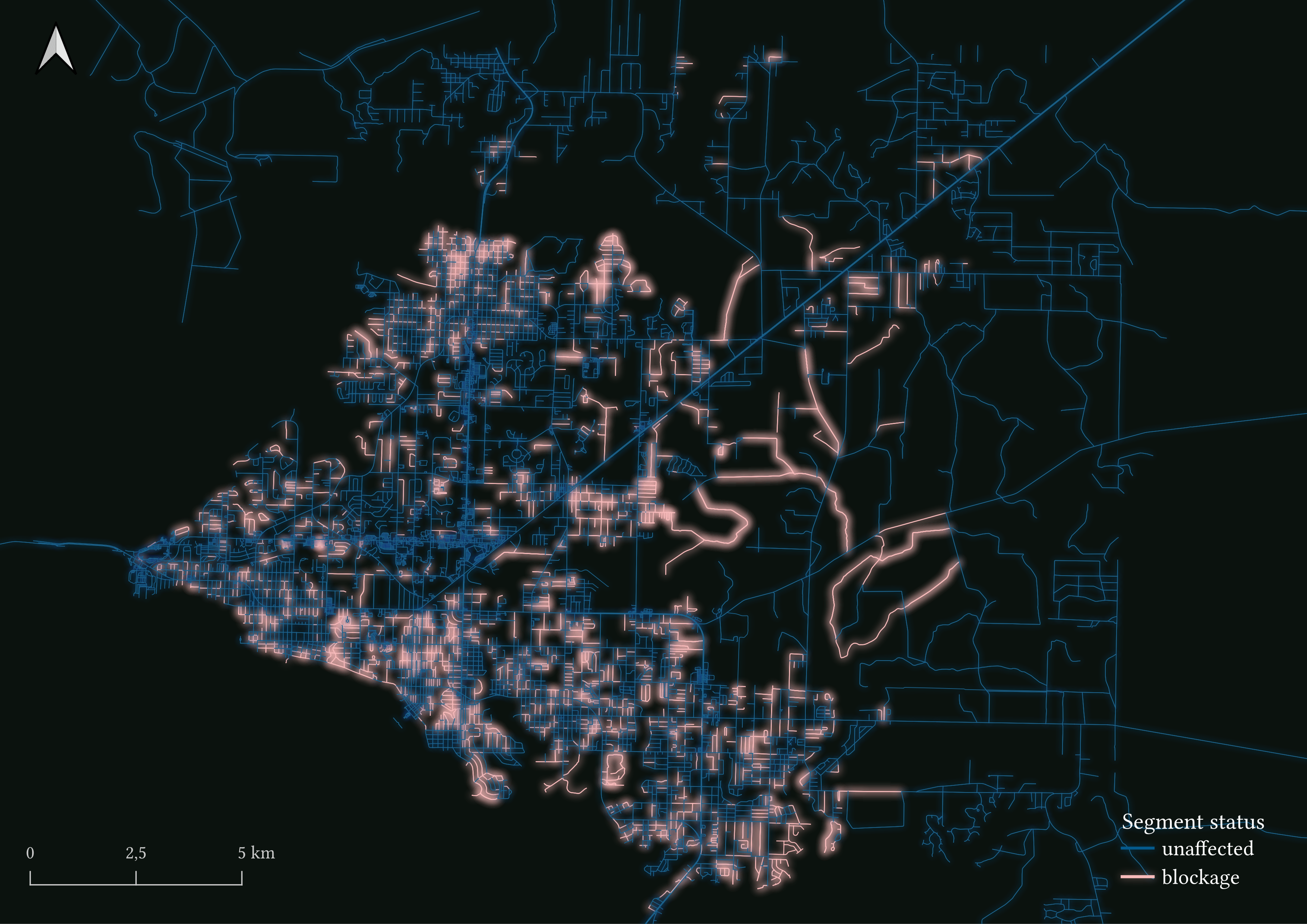}
    \caption{Panama City road transportation network directly affected by Hurricane Michael: segments blocked by fallen trees, debris, and water occurrence highlighted in brighter color. The road infrastructure experienced widespread damages, affecting multiple areas throughout the network. Original data derived from \autocite{noaa2018, osm}. Author's analysis and representation.}
    \label{fig:roadblockage}
\end{figure}

This study focuses on Panama City, Florida (approximately 37 thousand inhabitants), a community that experienced significant impacts causing disruptions to critical infrastructure like transportation systems and emergency services, such as hospitals, clinics, and fire stations. The storm left behind roads littered with debris and fallen trees---posing challenges for relief efforts following the event (See the calculation of obstacle occurrences on transportation networks in Figures \ref{fig:roadblockage}, \ref{fig:railblockage}).

\subsection{Transportation system}

For the analysis, six OSM highway classes (trunk, primary, secondary, tertiary, residential, and all other types considered together) were used (Table \ref{tab:roads}). In total, there are 14,982 street segments (edges) in the initial state of the system ($S_0$) in 2018; this number drops to 12,563 in the state affected by the hurricane ($S_1$) and rises to 16,978 in 2024 due to gradual recovery ($S_2$). In $S_0$, almost half of the roads are of residential type. This share decreased in the recovered state $S_2$. About 33 percent of roads belong to the class ``other'' in $S_0$ , and about 37 percent in $S_2$, while roads, such as trunk, primary, secondary, and tertiary account for about 2--7 percent in both $S_0$ and $S_2$. During the hurricane, residential roads were blocked the most (27.1 percent), followed by tertiary roads (6.5 percent) and other roads (6 percent). Only a few primary, secondary, and trunk roads were affected. Overall, the hurricane created barriers of various types on 16.1 percent of the roads. An increase in the number of road segments in the recovered system state $S_2$ is also observed, which may have altered the share distribution of roads of different classes. Since the residential streets were impacted the most, this portion of the transportation network can be classified as being particularly susceptible to damage.

\begin{table}[h]
    \footnotesize
    \caption{System description by the type of highway: a number and share of functioning edges before the hurricane impact in October 2018 ($S_0$), right after the impact ($S_1$), $b$---a number and share of segments blocked by the hurricane, and the number and share of edges in the year 2024 ($S_2$).  Author's calculation.}
    \centering
    \begin{tabular}{lllll|ll|ll}
    \toprule
       & $S_0$ (n) & $S_0$ (\%) & $S_1$ (n) & $S_1$ (\%) & $b$ (n) &  $b$ (\%) & $S_2$ (n) & $S_2$ (\%) \\
    \midrule
    Trunk & 455 & (3.0) & 454 &  (3.6) & 1 & (0.2) & 375 & (2.2) \\
    Primary & 327 & (2.2) & 318 &  (2.5) & 9 & (2.8) & 727 & (4.3) \\
    Secondary & 692 & (4.6) & 669 & (5.3) & 23 & (3.3) & 1093 & (6.4) \\
    Tertiary & 1023 & (6.8) & 956 & (7.6) & 67 & (6.5) & 986 & (5.8) \\
    Residential & 7436 & (49.6) & 5418  & (43.1) & 2018 & (27.1) & 7497 & (44.2) \\
    Other roads & 5049 & (33.7) & 4748 & (37.8) & 301 & (6.0) & 6300  & (37.1) \\
    Total & 14,982 & (100) & 12,563 & (100) & 2,419 & (16.1) & 16,978 & (100) \\
    \bottomrule
    \end{tabular}
    \label{tab:roads}
\end{table}

\subsection{Emergency system}

There are six types of critical services (See Table \ref{tab:services}). Each facility consists of a number of edges called \textit{facility edges} ($F_e$) to account for the possibility of reaching facilities from various sides. Together, these edges that are related to the same critical facility comprise a \textit{multiedge-facility} ($F_{me}$). According to OSM historical data, Panama City in 2018 ($S_0$) had 13 multiedge-facilities of type hospital, 6 multiedge clinics, 2 multiedge police stations, 8 multiedge fire stations, 3 multiedge pharmacies, and 2 multiedge shelters (Table \ref{tab:services}). The hurricane disabled 5 hospitals, 1 clinic, 2 fire stations, and 2 pharmacies. Despite observed disproportionate damage occurrence on facility-edges, some multiedge facilities remain functional due to a phenomenon the authors of this paper call \textit{hinder to access}, presuming that if not all streets around the facility are blocked, it is still possible to reach the building from an unaffected side. 
In $S_2$, a recovery of all services except one fire station is observed. Additionally, there has been an increase in $F_e$ number of hospitals, which can be related to the expansion of infrastructure of the present hospitals $F_{me}$. Therefore, hospitals, pharmacies, and fire stations were identified as the most at-risk locations during this event.

\begin{table}[h]
    \footnotesize
	\caption{Hurricane direct impact on critical infrastructure: number of facility edges $F_e$ (in parentheses---multiedge-facilities $F_{me}$). Comparison of the three system states as well as damage count $\delta$. Author's calculation.}
	\centering
		\begin{tabular}{l c c c c}
        \toprule
			& $S_0$ & $S_1$ & $\delta_1$ & $S_2$ \\
    	\midrule
        Hospital & 203 (\textbf{13}) & 170 (\textbf{8}) & 33 (\textbf{5}) & 308 (\textbf{13}) \\
        Clinic & 14 (\textbf{6}) & 12 (\textbf{5}) & 2 (\textbf{1}) & 17 (\textbf{6}) \\
        Police & 17 (\textbf{2}) & 16 (\textbf{2}) & 1 (\textbf{0}) & 18 (\textbf{2}) \\
        Fire station & 26 (\textbf{8}) & 18 (\textbf{6}) & 8 (\textbf{2}) & 22 (\textbf{7}) \\
        Shelter & 7 (\textbf{2}) & 5 (\textbf{2}) & 2 (\textbf{0}) & 9 (\textbf{2}) \\
        Pharmacy & 4 (\textbf{3}) & 2 (\textbf{1}) & 2 (\textbf{2}) & 4 (\textbf{3}) \\
        \bottomrule
    	\end{tabular}
	\label{tab:services}
\end{table}

\subsection{Social system}

The estimated total population within the network is 38,360 individuals, slightly higher than official statistics; this discrepancy arises because parts of other settlements that are organically connected to Panama City were included. The overall number of edges (street segments) in the studied network amounts to 14,982. Approximately 23.5 percent of the population was affected by the hurricane. These individuals reside on approximately 19.6 percent of all street segments (edges). Additionally, around 4 percent of the population---which is situated on 3.5 percent of edges---experienced indirect impacts from the storm; this means that obstacles did not occur directly on their streets but still resulted in access to services being obstructed due to blockages occurring nearby. In comparison to flood case in Jakarta where the same framework was applied \autocite{kiparisov2023}, here appears to be no significant difference in terms of access loss across various types of facilities; this may be due to differences in street organization and the pattern of damage caused by the hurricane, which was less localized and more uniformly distributed throughout the network. This finding also suggests that facilities within Panama City are located in a more dispersed manner, ensuring provision of essential services. Therefore, there is no observable difference in the susceptibility levels of specific types of facilities within the network.

\begin{table}[h]
    \footnotesize
    \caption{Approximate population and its share left without access to facilities in a result of indirect ($P_{ind}$) and both direct and indirect impact ($P_{t}$) from Hurricane Michael. Edge count and its share presented similarly for indirect ($n_{e_{ind}}$) and both direct and indirect impact ($n_{e_{t}}$). Author's calculation.}
    \centering
    \begin{tabular}{ l c c c c | c c c c}
    \midrule
        & $P_{ind}$ (n) & $P_{ind}$ (\%) &  $n_{e_{ind}}$ (n) & $n_{e_{ind}}$ (\%) & $P_{t}$ (n)  & $P_{t}$ (\%) & $n_{e_{t}}$ (n) &  $n_{e_{t}}$ (\%) \\
    \toprule
    Hospital   	    & 1548 & (4.04) & 518 & (3.5) & 9000 & (23.46) & 2937 & (19.6) \\
    Clinic 	        & 1545 & (4.03) &  516 & (3.4) & 8997 & (23.45) & 2935 & (19.59) \\
    Police station 	& 1548 & (4.04) &  518 & (3.5) & 9000 & (23.46) & 2937 & (19.6) \\
    Fire station 	& 1548 & (4.04) &  518 & (3.5) & 9000 & (23.46) & 2937 & (19.6) \\
    Pharmacy 	    & 1548 & (4.04) &  518 & (3.5) & 9000 & (23.46) & 2937 &  (19.6) \\
    Shelter 	    & 1542 & (4.01) &  516 & (3.4) & 8993 & (23.44) & 2935 & (19.59) \\
    \bottomrule
    \end{tabular}
    \label{tab:indirect}
\end{table}

According to Table \ref{tab:demographics}, it is evident that the population density for individuals aged 65 years old or more exceeds the average within the network. This difference is even more pronounced among male participants in this age group. The regression analysis demonstrates a strong relation between access loss and population density levels for male individuals aged 65 years old or more (See Table \ref{tab:olscurrent}); thus, this demographic is more exposed to the hurricane damage, which is consistent with prior research findings.

\begin{table}[h]
    \footnotesize
    \caption{Loss of access to services by mean population density aggregated by age and sex, including deviation from the system average to highlight more exposed to the hurricane groups of people. Author's calculation.}
    \centering
    \begin{tabular}{llllllll}
    \toprule
      & 0-14 (m) & 0-14 (f) & 15-64 (m) & 15-64 (f) & 65+ (m) & 65+ (f) \\
    \midrule
    Hospital & 61.458 & 59.88 & 211.361 & 223.411 & 49.824 & 64.882 \\
    Clinic & 61.269 & 59.752 & 211.048 & 223.014 & 49.909 & 64.975 \\
    Fire station & 61.458 & 59.88 & 211.361 & 223.411 & 49.824 & 64.882 \\
    Police station & 61.458 & 59.88 & 211.361 & 223.411 & 49.824 & 64.882 \\
    Pharmacy & 61.458 & 59.88 & 211.361 & 223.411 & 49.824 & 64.882 \\
    Shelter & 61.486 & 59.899 & 211.238 & 223.401 & 49.703 & 64.797 \\
    \midrule
    System average & 65.872 & 63.864 & 220.586 & 234.936 & 42.252 & 64.714 \\
    Deviation  & -4.386 & -3.965 & -9.225 & -11.525 & 7.657 & 0.261 \\
    \bottomrule
    \end{tabular}
    \label{tab:demographics}
\end{table}

\subsection{System robustness and resilience}

Analyzing the fundamental structure of the network and its system properties in three states---pre-hurricane ($S_0$), affected ($S_1$), and recovered ($S_2$)---it is possible to gain insights into its robustness and resilience. As shown in Table \ref{tab:netrobustness}, the hurricane has led to significant changes in the network's topology. The number of individual components increased to 1,057 ($\kappa$), while the maximum diameter ($D$) grew along with average shortest path length ($L$), global effective resistance ($R$), and average edge betweenness centrality ($\overline{b_e}$).

\begin{table}[h!]
    \footnotesize
	\caption{Road network robustness and resilience properties comparison between the three system states---the day before the hurricane impact in October 2018 ($S_0$), right after the impact ($S_1$), and in the year 2024 ($S_2$), where $n_v$ is the number of vertices, $n_e$ is the number of edges, $\kappa$---the number of components in the network, $d_{\max}$ is a diameter, $L$---average shortest path length, $E$ is a network efficiency, $C$---clustering coefficient, $R$ is the effective resistance, $b_e^{\max}$ indicates the maximum betweenness centrality, and $\overline{b_e}$ is average betweenness centrality. Author's calculation.}
	\centering
	\begin{tabular}{ l c c c c c c c c c  }
    \toprule
          & ${n_e}$ & ${\kappa}$ & $d_{\max}$ & $L$ & $E$ & $C$ & $R^{*}$ & $b_e^{\max}$ & $\overline{b_e}$ \\
	\midrule
	$S_0$ & $14,982$ & $1$     & 169 & 57.96 & 0.024 & 0.032 &  4.057        &  0.184	& 0.00387 \\
	$S_1$ & $12,577$ & $1,057$ & 194 & 66.36$^{**}$ & 0.016 & 0.027 &  8.01$^{**}$  &  0.14	&  0.00394 \\
	$S_2$ & $16,978$ & $1$     & 179 & 64.96 & 0.021 & 0.04  &  3.87         &  0.182	&  0.00383 \\
    \midrule
	$\delta_1$ & -16.1\%    & 105600\% &  14.8\% & 14.5\%  & -33.3\% & -15.6\%	& 97.4\%  & -23.9\%	 & 1.8\%	\\
	$\delta_2$ & 13.3\%  & 0\%   &  5.9\%   & 12.1\%  & -12.5\% & 25\%	& -4.6\%  & -1\%	  & -1\%	\\ 
    \bottomrule
	\end{tabular}
	\label{tab:netrobustness}
    \\
    \footnotesize{$^*$calculated for a fracture of the graph only.}
    \\
    \footnotesize{$^{**}$calculated for a giant connected component only.}
\end{table}

Conversely, the hurricane decreased network efficiency ($E$), clustering coefficient ($C$), and maximum edge betweenness centrality ($b_e^{\max}$). These changes are consistent with expectations. An increase in resistance is caused by a smaller amount of redundancies in the affected network, which implies a more dispersed or fragmented topology. Lower efficiency indicates fewer alternative routes for routing around potential bottlenecks or congestion points. A lower clustering coefficient suggests that neighboring nodes are less likely to be interconnected, resulting in a less compact regional structure within the network and indicating the reduction of redundancies. Finally, the decrease in maximum betweenness centrality indicates that some important edges were removed by the hurricane, which may be problematic for the system's resilience. Although an increase is not strongly accentuated and difficult to interpret, this may suggest that the system has become less decentralized, which is negative for its resilience. In the recovered state ($S_2$), mixed improvements and deteriorations across various network measures are observed. The average shortest path length ($L$) has increased compared to $S_0$, while network efficiency ($E$) has decreased, which are negative signs. The clustering coefficient ($C$) along with effective resistance ($R$), maximum edge betweenness centrality ($b_e^{\max}$), and average edge betweenness centrality ($\overline{b_e}$), however, have improved. These findings suggest that while some aspects of the network's structure may have been restored or enhanced, other elements remain vulnerable to future disruptions.

\begin{figure}[h]
    \centering
    \includegraphics[width=.327\textwidth]{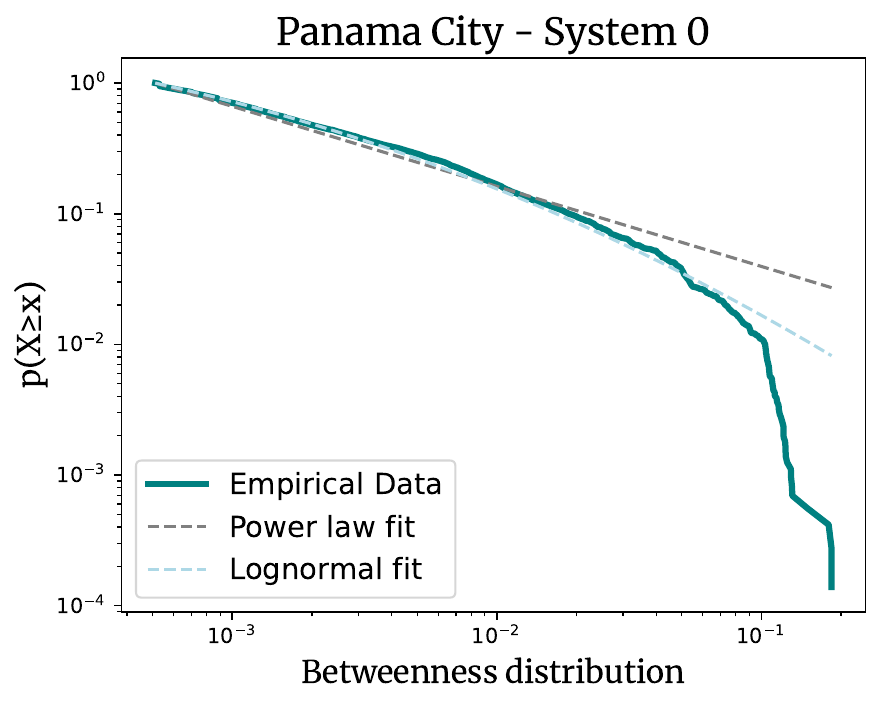}
    \includegraphics[width=.327\textwidth]{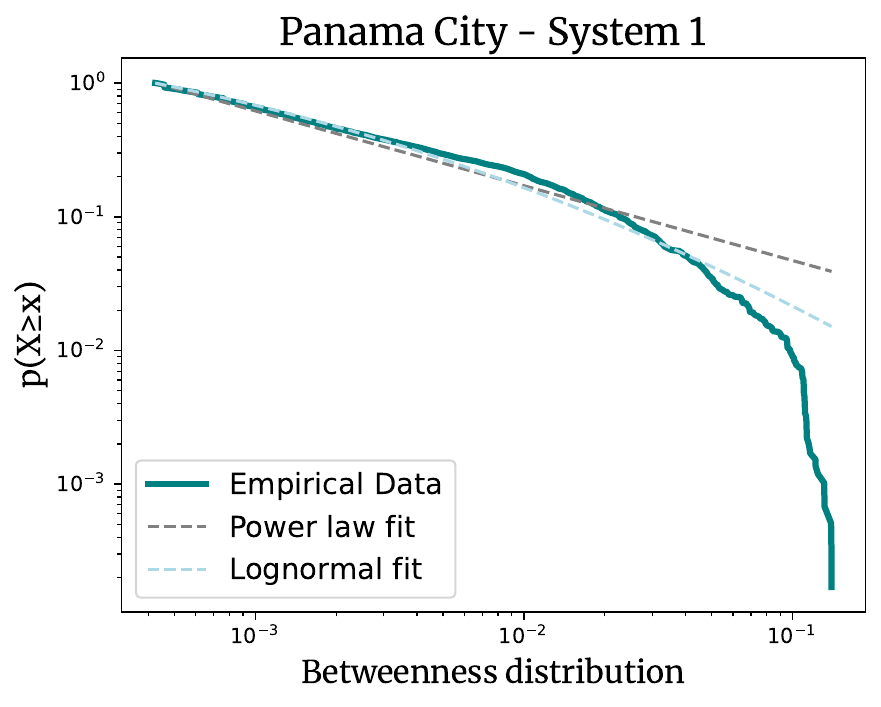}
    \includegraphics[width=.327\textwidth]{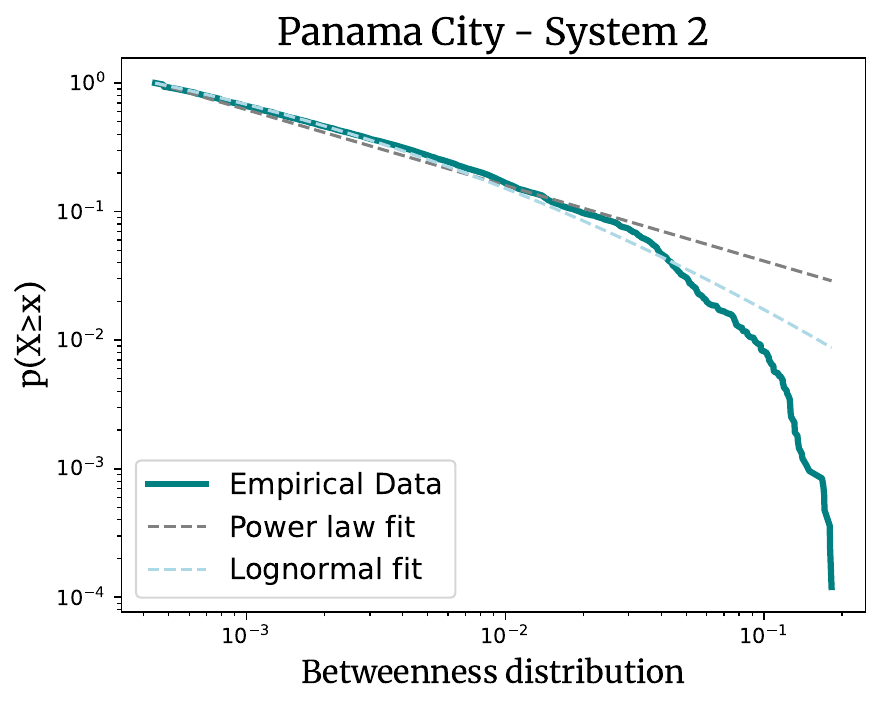}
    \centering
    \caption{Betweenness centrality power law distribution. Author's calculation.}
    \label{fig:bcdistribution}
\end{figure}

Changes in the distribution function of betweenness centrality parameters across three systems were also analyzed (as depicted in Figure \ref{fig:bcdistribution}). In a power-law function, $\alpha$ (alpha) represents the exponent or scaling factor that determines how steeply the function increases or decreases with respect to $x$. $x_{\min}$ signifies the value of $x$ at which the function reaches its minimum point. A decrease in $\alpha$ is observed within the affected system and an increase following recovery---for three states of the system $S_0$, $S_1$, and $S_2$ the values of $\alpha$ are 1.61, 1.56, and 1.59 respectively. $x_{\min}$ in $S_0$ is equal to 0.00051, it falls to 0.00042 in $S_1$, and grows to 0.00044 in  $S_2$. A smaller $\alpha$ and a more uniform distribution of $x$ in the tail of the recovered system compared to its pre-hurricane state suggest some positive improvements within the transportation network following recovery efforts.

The analysis of the railroad system reveals that all resilience metrics have decreased following damage caused by the hurricane (Table \ref{tab:railways}); therefore, it is essential to note that these properties may not be generalizable across different transportation networks due to variations in size and structure among studied graphs---the railway system was completely disabled by the hurricane (Figure \ref{fig:railblockage}), whereas the street network managed to maintain a significant degree of connectivity throughout the event.


\subsection{Vulnerability assessment through simulation and treatment scenario}

\begin{figure}[h]
    \includegraphics[width=1\textwidth]{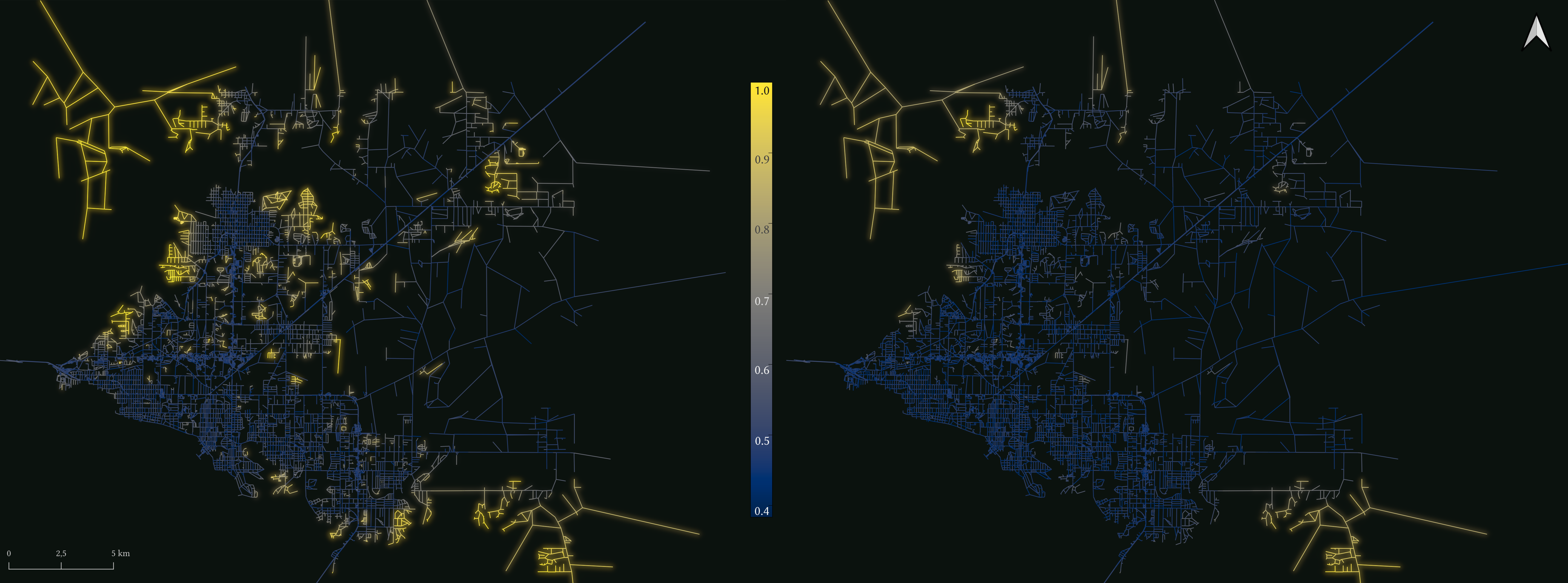}
    \centering
    \caption{The presented maps highlight the susceptibility of Panama City to a loss of access to at least one critical service by its population due to potential natural disasters such as Hurricane Michael or stronger events. On the left-hand side: a map illustrates the final ranking of the proposed probabilistic model of susceptibility to losing access that follows base patterns of Hurricane Michael with a certain likelihood of exceeding its destructive powers calculated for each street segment. On the right-hand side: another map depicts the vulnerability index of street segments based on these same patterns but introducing treatment scenario for residential streets, including increasing the distance of the trees from the road and introducing more hurricane-prone species. One can see how altering just a single simulated variable can significantly enhance the overall state of affairs.}
    \label{fig:vulnerability}
\end{figure}

\begin{figure}[h!]
    \centering
    \includegraphics[width=.54673\textwidth]{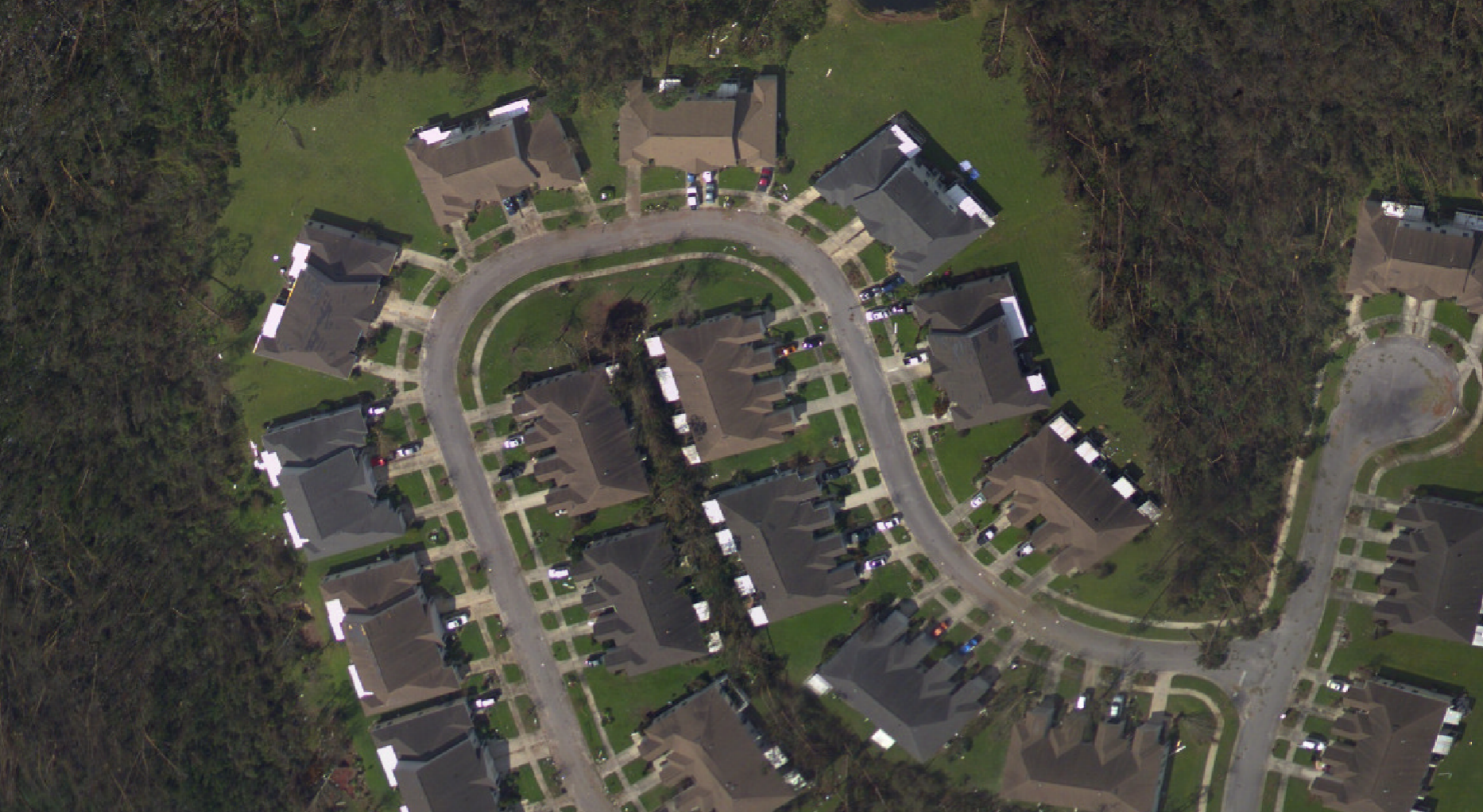}
    \includegraphics[width=.44\textwidth]{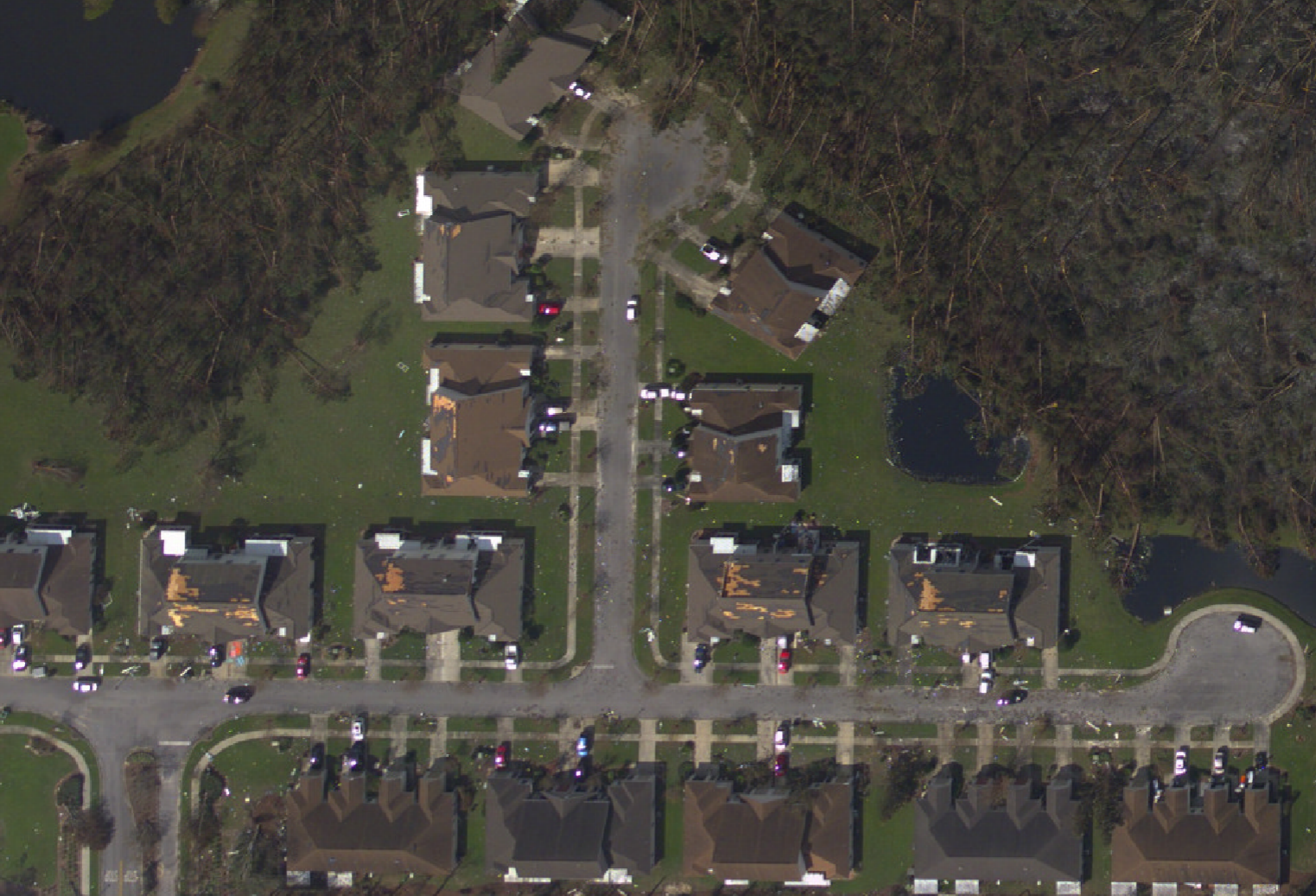}
    \caption{Examples of improved residential planning involve strategically positioning trees at a secure distance from roadways and buildings to minimize potential harm during disasters. The images illustrate a neighborhood that was less impacted by tree blockage due to maintaining a safe distance for the trees planted away from the roads. Source: \autocite{noaa2018}.}
    \label{fig:betterplanning}
\end{figure}

A key aspect of this research involves conducting simulations on the post-recovery street network $S_2$, with the aim of identifying general vulnerabilities as discussed in Section \ref{sec:simulation}. The results of these simulations are depicted in Figure \ref{fig:vulnerability}, which reveals that segments located at the peripheries of Panama city were found to be particularly susceptible to access loss. During the preparatory stage while authors studied satellite images, it was noticed that some residential streets were less affected by obstructions caused by Hurricane Michael. Upon closer inspection, these less impacted segments of roads maintained a greater distance from trees (Figure \ref{fig:betterplanning}). This led the authors to the idea of running simulations to show how system-wide maintenance of tree setbacks from roads and planting more robust tree species, as suggested by \cite{ward2017}, could improve system resilience.

A significant improvement is observed in the simulation scenario with a treatment, when stricter regulations for tree planting along residential streets are implemented (Figure \ref{fig:vulnerability}); this suggests that proactive measures can play an essential role in decreasing network vulnerability and that if this urban design is maintained across the board, another hurricane would result in fewer blocked roadways due to fallen trees and, consequently, fewer people losing access to critical services. After introduction, the treatment scenario significantly reduces the global vulnerability index ($V$), lowering it from 0.56 to 0.32. The level of vulnerability experienced by different demographic groups within the city is considered (Tables \ref{tab:olssim1}). It is apparent that male individuals aged 65 years old or more are disproportionately affected across all eventualities; this trend is observed regardless of whether treatment interventions are implemented or not. The same was observed in the actual event of Hurricane Michael.

\subsection{Way forward}

In general, adding a state related to recovery has taken this research one step closer towards a comprehensive assessment of urban resilience dynamics. Future investigations should explore additional aspects such as urban form considerations and facility placement. Obtaining higher quality data and conducting ground surveys will undoubtedly yield better outcomes. A thorough examination of the destruction caused by a natural disaster would not be possible without access to high-resolution satellite imagery, which is not always available; therefore, it is essential for space agencies and private companies to make these data more accessible. Combining this quantitative approach with qualitative insights from sociological surveys and interviews may present an interesting avenue for further methodological development. The impact of seasonality on tourist visits should also be considered in future studies.

Network analysis can be enhanced by introduction of configuration models as the basis for comparison. Comparison to other similar cities unaffected by hurricane will also add to the analysis. More vital services can be added in future research, e.g. following the US Department of Homeland Security's (FEMA) concept of \textit{community lifelines}, which represent critical services in a community that enable society to function stably. These lifelines include such categories as safety and security, food, hydration, and shelter, health and medical, energy, communications, hazardous materials, and water systems  \autocite{fema2024}. Comparing patterns of various types of catastrophes that damage or obstruct transportation networks and understanding their impact on city resilience is particularly important given the context of climate change and population growth, especially in areas prone to disasters.

\section{Conclusion}
\label{conclusion}

The effects of Hurricane Michael on Panama City and its surrounding areas have been profoundly impactful, and this study aimed to contribute towards understanding these consequences better, with the ultimate goal of informing more resilient infrastructure networks for future disaster events.
The estimated total population within the network was 38,360 individuals, with approximately 23.5 percent affected by the hurricane and residing on around 19.6 percent of all street segments. Additionally, about 4 percent of the population experienced indirect impacts from the storm due to blockages occurring nearby. The transportation system analysis reveals that residential roads were most affected, with 27.1 percent of them being blocked. Tertiary and ``other'' roads also experienced significant disruptions. Hospitals, pharmacies, and fire stations were the most at-risk facilities during Hurricane Michael. Despite significant damage to these facilities' edges, some multiedge structures remained functional due to access from unaffected sides. In the recovered state, all services except one fire station experienced restoration, with an increase in hospital facility edge numbers attributed to infrastructure expansion.

Studying the fundamental structure of a network affected by a disaster provides valuable insights into how real-world catastrophes impact measures typically associated with graph robustness and resilience. This research demonstrates these effects are consistent with existing literature, suggesting their applicability in studying the consequences of hurricanes on road networks specifically. Hurricane damage caused significant changes in road network topology: efficiency, clustering coefficient, and maximum edge betweenness centrality decreased after the hurricane, mixed improvements and deteriorations were observed across various measures in the recovered state, and a smaller exponent of betweenness centrality distribution indicates positive improvements following recovery efforts. Simulations on post-recovery street networks identified general vulnerabilities and showed that proactive measures play an essential role in enhancing network resilience. The analysis demonstrated an existing relation between access loss and population density levels for males aged 65 years old or more; thus, this demographic is classified as being more exposed to the hurricane damage, which is consistent with prior research findings. The simulations have also confirmed the pattern of vulnerability for these age groups.

Usually the challenges encountered arise from the systems people designed \autocite{sterman2002}. Living near the coast increases one's vulnerability to natural disasters; people, however, continue to put themselves at risk by living in areas prone to catastrophe inherently inviting trouble by standing in the way of natural events. As previously mentioned in the Section \ref{background}, there are still many reasons why individuals choose to reside in coastal regions despite their susceptibility to hurricanes and other natural hazards. A notable example, even after being severely damaged by Hurricane Michael, places like Mexico Beach are being repopulated rather than remaining solely as tourist destinations \autocite{wfsu2023}. Although there are already some instances of moving or abandonment of entire towns located in high-risk disaster areas \autocite{bbc2024}, mass relocation of affected cities is not feasible, and the focus should remain on improving the resilience of at-risk communities. The only means by which the susceptibilities of systems can be reduced is through a deeper comprehension of their nature, followed by improvements in their future design. This study has demonstrated that introducing a system-wide improvement to a particularly vulnerable segment can enhance the overall system's performance.

\section*{Acknowledgement(s)}
The authors are grateful for the insightful advice from Georg Pflug, engaging mathematical discussions with Andrey Krasovskiy, and support provided by Augusto Jr Salvador during updates to the literature review. This paper has greatly benefited from comments made by Márton Pósfai and Trevor Covington who read the earlier version of the entire manuscript. 

\section*{Disclosure statement}
The authors report there are no competing interests to declare.

\section*{Data availability statement}
The data that support the findings of this study are openly available in zenodo at https://doi.org/10.5281/zenodo.13932741, reference number \cite{hurricanenet}.

\section*{Author contributions}

Conceptualization, P.O.K.; Data curation, P.O.K.; Formal analysis, P.O.K.; Investigation, P.O.K.; Methodology, P.O.K.; Project administration, V.V.L.; Resources, V.V.L.; Software, P.O.K.; Supervision, V.V.L.; Validation, P.O.K.; Visualization, P.O.K.; Writing-original draft, P.O.K.; Writing-review \& editing, P.O.K.


\printbibliography

\newpage
\appendix
\label{sec:appendix}
\section{Supplemental materials to ``Integrated GIS- and network-based framework for assessing urban critical infrastructure accessibility and resilience: the case of Hurricane Michael''}

\begin{figure}[h!]
    \centering
    \includegraphics[width=0.9\textwidth]{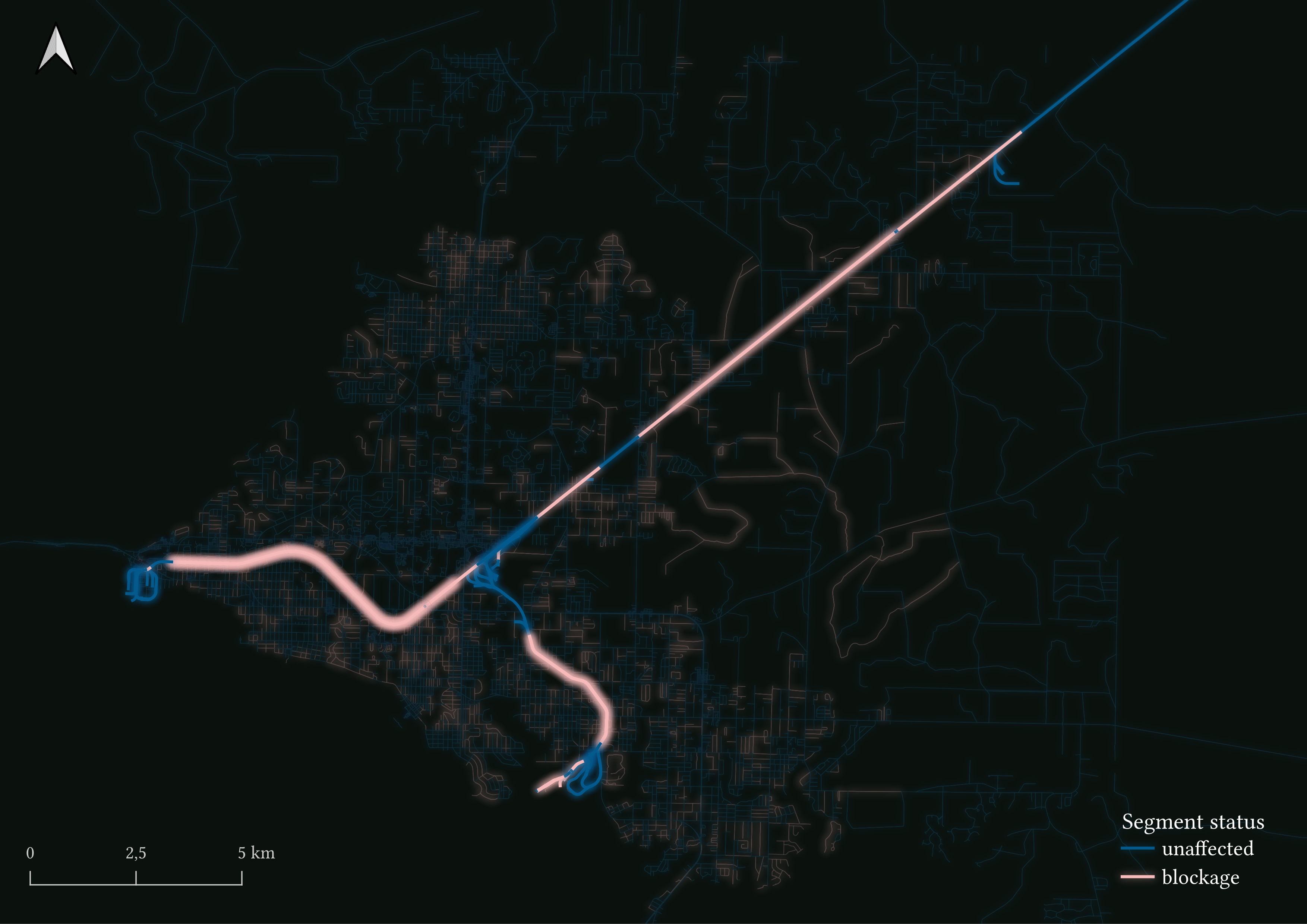}
    \caption{Panama City railway network directly affected by Hurricane Michael: segments blocked by fallen trees, debris, and water occurrence highlighted in brighter color. The rail transportation system suffered severe disruptions due to its linear design and limited branches. Original data derived from \cite{noaa2018, osm}. Author's analysis and representation.}
    \label{fig:railblockage}
\end{figure}

\begin{figure}[H]
    \centering
    \includegraphics[width=.327\textwidth]{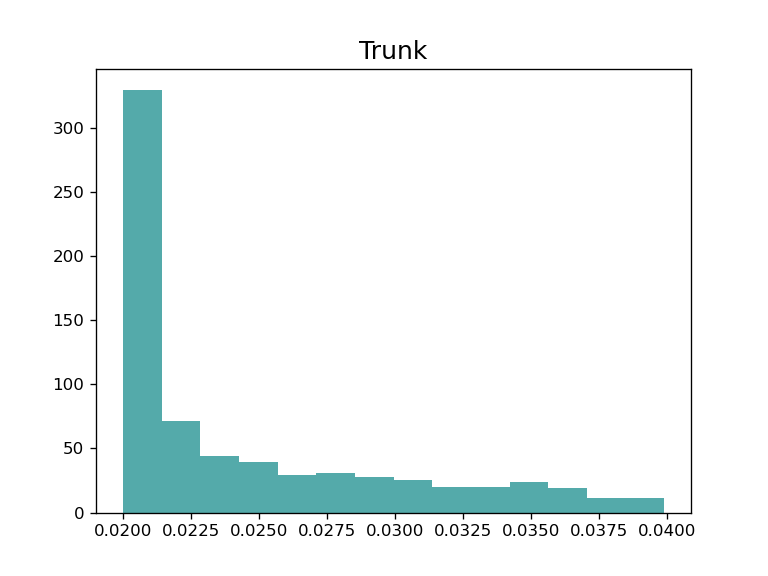}
    \includegraphics[width=.327\textwidth]{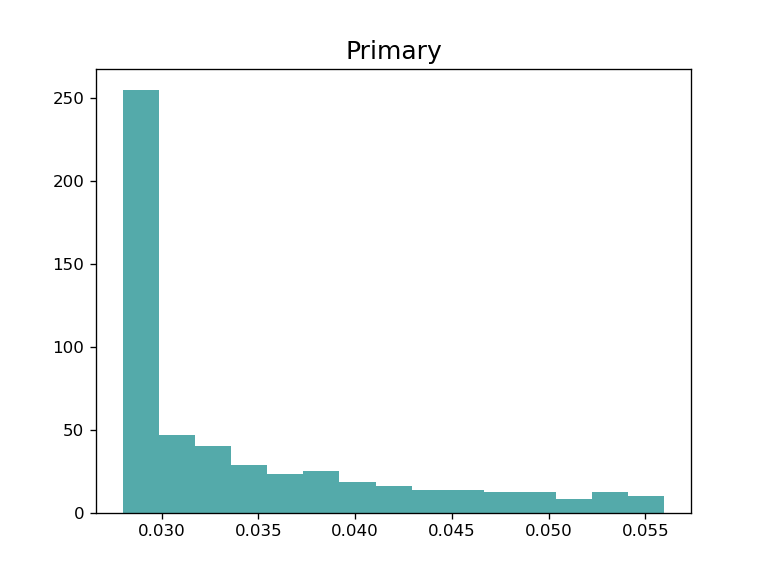}
    \includegraphics[width=.327\textwidth]{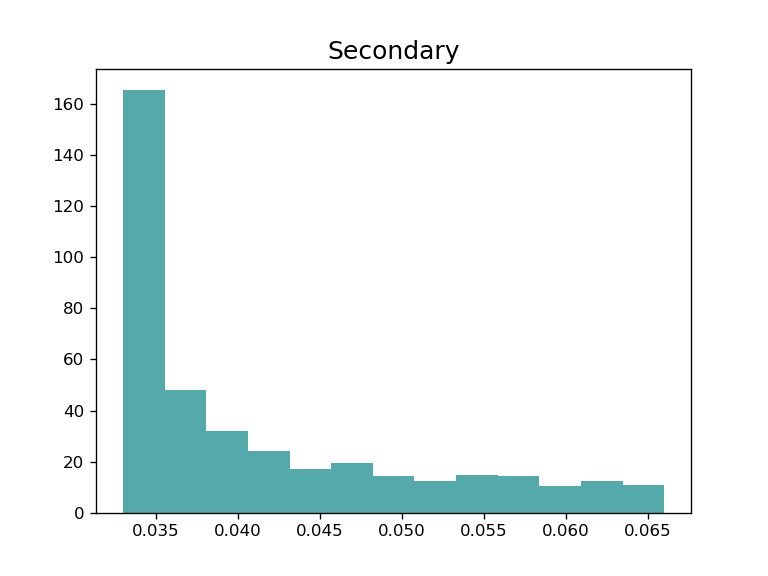}
    \includegraphics[width=.327\textwidth]{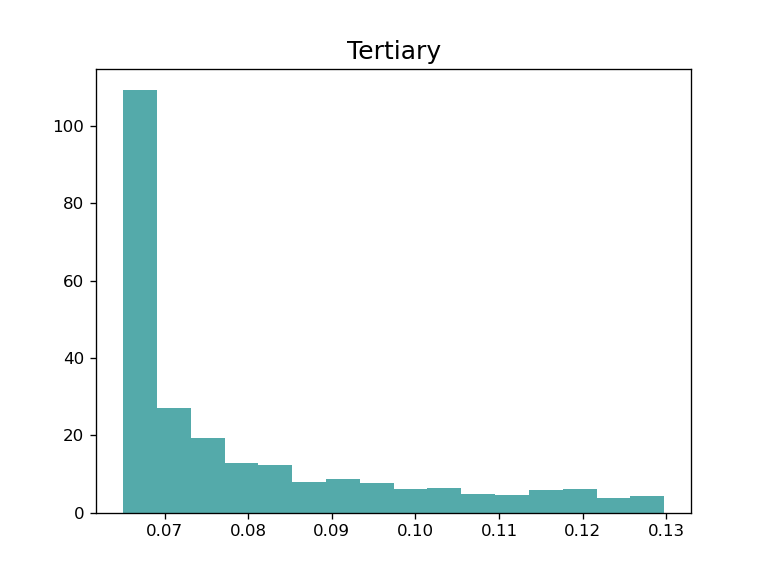}
    \includegraphics[width=.327\textwidth]{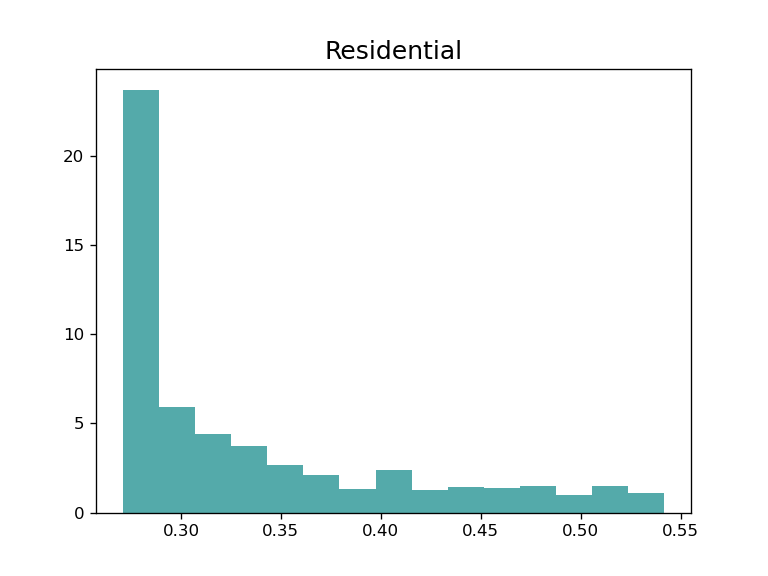}
    \includegraphics[width=.327\textwidth]{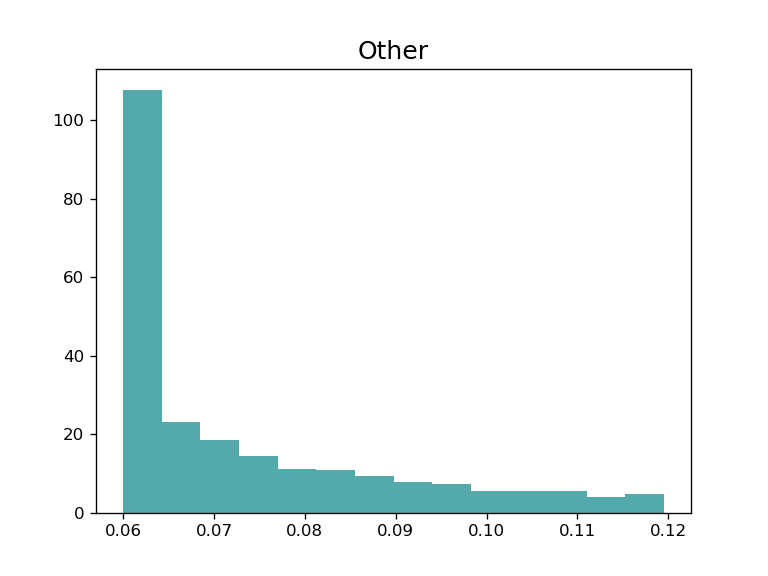}
    \centering
    \caption{Sample probability density functions randomly generated for various street types.}
    \label{fig:pdfsample}
\end{figure}

\begin{table}[H]
    \footnotesize
	\caption{Railway network robustness and resilience properties comparison between the three system states---the day before the hurricane impact in October 2018 ($S_0$), right after the impact ($S_1$), and in the year 2024 ($S_2$), where $n_e$ is the number of edges, $\kappa$---the number of components in the network, $d_{\max}$ is a diameter, $L$---average shortest path length, $E$ is a network efficiency, $C$---clustering coefficient, $R$ is the effective resistance, $b_e^{\max}$ indicates the maximum betweenness centrality, and $\overline{b_e}$ is average betweenness centrality. Author's calculation.}
	\centering
    \begin{tabular}{c c c c c c c c c c}
    \toprule
            & $n_e$ & $\kappa$ & $d_{\max}$ & $L$ & $E$ & $C$ & $R$ & $b_e^{\max}$ & $\overline{b_e}$ \\
	\midrule
	$S_0$ & $250$ & $1$  & 36 & 15.57 & 0.103 & 0.046 &  11.35   & 0.487 & 0.062 \\
	$S_1$ & $215$ & $29$ & 17 & 7.22 & 0.046  & 0.031 &   4.72  & 0.029 & 0.007 \\
    \hline
	$\delta_1$ & -14\% &  2800\% & -52.8\% & -53.6\% & -55\%  & -32.6\%  & -58.4\%	& -94\% & -88.7\%   \\
    \bottomrule
	\end{tabular}
	\label{tab:railways}
\end{table}

\begin{table}[h!]
\footnotesize
\label{tab:olscurrent}
    \begin{center}
    \caption{OLS Regression Results---Hurricane Michael}
    \begin{tabular}{lclc}
    \toprule
    \textbf{Dep. Variable:}    &     \textbf{loss of access}     & \textbf{  R-squared:         } &     0.036   \\
    \textbf{Model:}            &       OLS        & \textbf{  Adj. R-squared:    } &     0.036   \\
    \textbf{Method:}           &  Least Squares   & \textbf{  F-statistic:       } &     94.15   \\
    \textbf{No. Observations:} &       14982      & \textbf{  Prob (F-statistic):} & 1.44e-116   \\
    \textbf{Df Residuals:}     &       14975      & \textbf{  Log-Likelihood:    } &   -7142.2   \\
    \textbf{Df Model:}         &           6      & \textbf{  AIC:               } & 1.430e+04   \\
    \textbf{Df Residuals:}     &       14975      & \textbf{  BIC:               } & 1.435e+04   \\
    \bottomrule
    \end{tabular}
    \begin{tabular}{lcccccc}
                   & \textbf{coef} & \textbf{std err} & \textbf{t} & \textbf{P$> |$t$|$} & \textbf{[0.025} & \textbf{0.975]}  \\
    \midrule
    \textbf{const} &       0.1113  &        0.006     &    18.314  &         0.000        &        0.099    &        0.123     \\
    \textbf{male 0-14}    &       0.0006  &        0.000     &     1.995  &         0.046        &     9.74e-06    &        0.001     \\
    \textbf{female 0-14}    &      -0.0003  &        0.000     &    -1.004  &         0.315        &       -0.001    &        0.000     \\
    \textbf{male 15-65}   &       0.0001  &        0.000     &     1.441  &         0.150        &    -5.27e-05    &        0.000     \\
    \textbf{female 15-65}   &      -0.0003  &        0.000     &    -2.358  &         0.018        &       -0.001    &    -5.26e-05     \\
    \textbf{male 65+}   &       0.0047  &        0.000     &    17.751  &         0.000        &        0.004    &        0.005     \\
    \textbf{female 65+}   &      -0.0014  &        0.000     &   -12.340  &         0.000        &       -0.002    &       -0.001     \\
    \bottomrule
    \end{tabular}
    \end{center}
\end{table}

\begin{table}[h!]
\footnotesize
\label{tab:olssim1}
    \begin{center}
    \caption{OLS Regression Results---Simulation following destructive patterns of Hurricane Michael with a certain likelihood of exceeding its damage.}
    \begin{tabular}{lclc}
    \toprule
    \textbf{Dep. Variable:}    &    \textbf{loss of access}     & \textbf{  R-squared:         } &     0.084   \\
    \textbf{Model:}            &       OLS        & \textbf{  Adj. R-squared:    } &     0.084   \\
    \textbf{Method:}           &  Least Squares   & \textbf{  F-statistic:       } &     260.3   \\
    \textbf{No. Observations:} &       16978      & \textbf{  Prob (F-statistic):} & 1.09e-319   \\
    \textbf{Df Residuals:}     &       16971      & \textbf{  Log-Likelihood:    } &    11010.   \\
    \textbf{Df Model:}         &           6      & \textbf{  AIC:               } & -2.201e+04  \\
    \textbf{Covariance Type:}  &    nonrobust     & \textbf{  BIC:               } & -2.195e+04  \\
    \bottomrule
    \end{tabular}
    \begin{tabular}{lcccccc}
                   & \textbf{coef} & \textbf{std err} & \textbf{t} & \textbf{P$> |$t$|$} & \textbf{[0.025} & \textbf{0.975]}  \\
    \midrule
    \textbf{const} &       0.6010  &        0.002     &   324.965  &         0.000        &        0.597    &        0.605     \\
    \textbf{male 0-14}    &   -3.064e-05  &     5.96e-05     &    -0.514  &         0.607        &       -0.000    &     8.62e-05     \\
    \textbf{female 0-14}    &    3.995e-06  &     6.93e-05     &     0.058  &         0.954        &       -0.000    &        0.000     \\
    \textbf{male 15-65}   &   -7.429e-05  &     2.97e-05     &    -2.505  &         0.012        &       -0.000    &    -1.62e-05     \\
    \textbf{female 15-65}   &      -0.0001  &     3.79e-05     &    -3.486  &         0.000        &       -0.000    &    -5.78e-05     \\
    \textbf{male 65+}   &       0.0017  &     7.92e-05     &    22.022  &         0.000        &        0.002    &        0.002     \\
    \textbf{female 65+}   &      -0.0010  &     3.35e-05     &   -30.903  &         0.000        &       -0.001    &       -0.001     \\
    \bottomrule
    \end{tabular}
    \end{center}
\end{table}

\begin{table}[h!]
\footnotesize
\label{tab:olssim2}
    \begin{center}
    \caption{OLS Regression Results---Simulation taking the treatment scenario on residential streets in consideration.}
    \begin{tabular}{lclc}
    \toprule
    \textbf{Dep. Variable:}    &    \textbf{loss of access}     & \textbf{  R-squared:         } &     0.060   \\
    \textbf{Model:}            &       OLS        & \textbf{  Adj. R-squared:    } &     0.060   \\
    \textbf{Method:}           &  Least Squares   & \textbf{  F-statistic:       } &     180.4   \\
    \textbf{No. Observations:} &       16978      & \textbf{  Prob (F-statistic):} & 1.75e-223   \\
    \textbf{Df Residuals:}     &       16971      & \textbf{  Log-Likelihood:    } &    19097.   \\
    \textbf{Df Model:}         &           6      & \textbf{  AIC:               } & -3.818e+04  \\
    \textbf{Covariance Type:}  &    nonrobust     & \textbf{  BIC:               } & -3.813e+04  \\
    \bottomrule
    \end{tabular}
    \begin{tabular}{lcccccc}
                   & \textbf{coef} & \textbf{std err} & \textbf{t} & \textbf{P$> |$t$|$} & \textbf{[0.025} & \textbf{0.975]}  \\
    \midrule
    \textbf{const} &       0.3450  &        0.001     &   300.379  &         0.000        &        0.343    &        0.347     \\
    \textbf{male 0-14}    &   -9.775e-05  &      3.7e-05     &    -2.641  &         0.008        &       -0.000    &    -2.52e-05     \\
    \textbf{female 0-14}    &   -7.712e-05  &     4.31e-05     &    -1.791  &         0.073        &       -0.000    &     7.27e-06     \\
    \textbf{male 15-65}   &   -9.864e-05  &     1.84e-05     &    -5.356  &         0.000        &       -0.000    &    -6.25e-05     \\
    \textbf{female 15-65}   &    3.902e-05  &     2.35e-05     &     1.659  &         0.097        &    -7.08e-06    &     8.51e-05     \\
    \textbf{male 65+}   &       0.0004  &     4.92e-05     &     7.494  &         0.000        &        0.000    &        0.000     \\
    \textbf{female 65+}   &      -0.0003  &     2.08e-05     &   -16.254  &         0.000        &       -0.000    &       -0.000     \\
    \bottomrule
    \end{tabular}
    \end{center}
\end{table}

\end{document}